\documentclass[11pt,a4paper]{article}
\pdfoutput=1

\usepackage{jheppub}
\usepackage[utf8]{inputenc}
\usepackage{enumitem}
\newcommand{\D}{\mathrm{d}}
\newcommand{\Ocal}{\mathcal{O}}
\newcommand{\Acal}{\mathcal{A}}
\newcommand{\Bcal}{\mathcal{B}}

\newcommand{\Ccal}{\mathcal{C}}

\newcommand{\AdS}{\mathrm{AdS}}
\newcommand{\be}{\begin{equation}}
\newcommand{\ee}{\end{equation}}
\newcommand{\bea}{\begin{eqnarray}}
\newcommand{\eea}{\end{eqnarray}}
\newcommand{\beas}{\begin{eqnarray*}}
\newcommand{\eeas}{\end{eqnarray*}}
\newcommand{\ba}{\begin{array}}
\newcommand{\ea}{\end{array}}
\newcommand{\tr}{{\rm tr}}

\newcommand{\bracket}[3]{\langle #1 \vert #2 \vert #3 \rangle}
\newcommand{\avg}[1]{\langle #1 \rangle} 

\newcommand{\uavg}[1]{\langle #1 \rangle_{\mathrm{UHP}}^b}
\newcommand{\vblock}[6]{\mathcal{F}(c, #1 ; [#2,#3,#4,#5] | #6)}

\usepackage[normalem]{ulem}
\usepackage{etoolbox}
\usepackage{cancel}
\usepackage{todonotes}

\newtoggle{editing}
\togglefalse{editing}

\iftoggle{editing}{%
	\setlength{\marginparwidth}{2cm}
	\usepackage[draft]{showlabels}
	
	\newcommand{\Mark}[1]{\textbf{\color{blue}[#1  --\textit{Mark}]}}
	\newcommand{\jamie}[1]{\textbf{\color{red}[#1  --\textit{Jamie}]}}
	\newcommand{\david}[1]{\textbf{\color{purple}[#1  --\textit{David}]}}
}{%
	\newcommand{\Mark}[1]{\textbf{\color{blue}[]}}
	\newcommand{\jamie}[1]{\textbf{\color{red}[]}}
	\newcommand{\david}[1]{\textbf{\color{purple}[]}}
}

\title{
BCFT entanglement entropy 
at large central charge 
and the black hole interior
}

\author[a]{James Sully,}
\author[a]{Mark Van Raamsdonk,}
\author[a]{and David Wakeham}
\date{\today}

\affiliation[a]{Department of Physics and Astronomy,\\
University of British Columbia, %6224 Agricultural Road,
Vancouver,
BC V6T 0C2, Canada}

\emailAdd{sully@phas.ubc.ca}
\emailAdd{mav@phas.ubc.ca}
\emailAdd{daw@phas.ubc.ca}
\abstract{
In this note, we consider entanglement and Renyi entropies for spatial subsystems of a boundary conformal field theory (BCFT) or of a CFT in a state constructed using a Euclidean BCFT path integral. Holographic calculations suggest that these entropies undergo phase transitions as a function of time or parameters describing the subsystem; these arise from a change in topology of the RT surface. In recent applications to black hole physics, such transitions have been seen to govern whether or not the bulk entanglement wedge of a (B)CFT region includes a portion of the black hole interior and have played a crucial role in understanding the semiclassical origin of the Page curve for evaporating black holes.

In this paper, we reproduce these holographic results via direct (B)CFT calculations. Using the replica method, the entropies are related to correlation functions of twist operators in a Euclidean BCFT. These correlations functions can be expanded in various channels involving intermediate bulk or boundary operators. Under certain sparseness conditions on the spectrum and OPE coefficients of bulk and boundary operators, we show that the twist correlators are dominated by the vacuum block in a single channel, with the relevant channel depending on the position of the twists. These transitions between channels lead to the holographically observed phase transitions in entropies.}

\begin{document}
\maketitle

\makeatletter
\@starttoc{toc}
\makeatother
\newpage
\section{Introduction}
\label{sec:intro}

In this note, we discuss the direct CFT calculation of entanglement and R\'{e}nyi entropies for an interval (or collection of intervals) in some related 1+1-dimensional systems:
\begin{enumerate}
\item
The vacuum state of a boundary conformal field theory (BCFT) on a half-space.
\item
The state of a CFT on a circle produced by a Euclidean path integral with a boundary in the Euclidean past.
\item
A pair of BCFTs in a thermofield double state.
\end{enumerate}
For holographic examples of these systems, calculations making use of the the Ryu-Takayanagi (RT) formula \cite{Ryu2006a} suggest that the entanglement entropy can undergo phase transitions related to a change in topology of the RT surface (for examples, see Figures 1, 4 and 5). In this paper, we show that these holographic results can be reproduced with direct BCFT calculations by assuming large central charge and certain conditions on the CFT spectrum and OPE data. In particular, we show that the assumption of {\it vacuum block dominance} for BCFT correlators of twist operators is equivalent to a simple holographic prescription \cite{Karch2000,Takayanagi2011} for the duals of BCFTs where the CFT boundary extends into the bulk as a purely gravitational end-of-the-world (ETW) brane with tension related to the boundary entropy of the BCFT (as suggested by \cite{Takayanagi2011}).\footnote{ 
The philosophy of reproducing gravitational results from vacuum block dominance goes by the amusing name ``It from Id" \cite{Anous2016}.}

This simple holographic prescription with a purely gravitational ETW brane is not expected to be valid universally for holographic BCFTs. In many cases, we can have non-vanishing one-point functions for light scalar operators in the BCFT. These translate to back-reacting scalar fields in the dual geometry that we can think of as being sourced by the ETW brane. In these cases with back-reacting scalars, the entanglement entropy for an interval in the phase corresponding to a connected RT surface  (e.g. Figure 1, top left) is a complicated function of the interval size and location and is not expected to be reproduced by the vacuum contribution in some channel.

On the other hand, the entanglement entropy in the phase corresponding to a disconnected RT surface is still simple and universal (see equation (\ref{disc})) and reproduced by the vacuum block contribution to the boundary channel. Thus, while vacuum block dominance is not expected to hold universally for holographic BCFTs, the gravity results indicate that it does hold in this phase. This suggests a particular sparseness condition (\ref{Bbound}) that should hold for the spectrum of any holographic BCFT.

\subsubsection*{Probing black hole interiors}

Our results have various applications to the physics of black hole interiors:

\begin{itemize}
    \item 
In \cite{Cooper2018}, it was argued that CFT states of type 2 above correspond to black hole microstates with a specific behind-the-horizon region whose geometry can be deduced by a standard holographic gravity calculation. Holographic calculations using the Ryu-Takayanagi (RT) formula \cite{Ryu2006a} suggested that in many cases, the black hole interior can be probed via the entanglement entropy of sufficiently large subsystems of the CFT. Our direct CFT calculations confirm this expectation, precisely matching the results of \cite{Cooper2018}.
\item 
In \cite{Rozali2019}, a pair of BCFTs in the thermofield double state was proposed as a microscopic model of a black hole in equilibrium with its Hawking radiation, following \cite{Penington:2019npb,Almheiri2019b,Almheiri2019a}. The RT surface for a fixed portion of the radiation system exhibits a phase transition as a function of time as the black hole and radiation system interact. After the transition, the entanglement wedge of this subsystem includes a portion of the black hole interior, suggesting that the black hole has transferred information about its interior to the radiation system through interaction. Our CFT calculations directly confirm the behavior of entanglement entropy surmised from the holographic calculation.
\end{itemize}

\subsubsection*{BCFT methods}

\begin{figure}[t]
  \centering
  \includegraphics[scale=0.2]{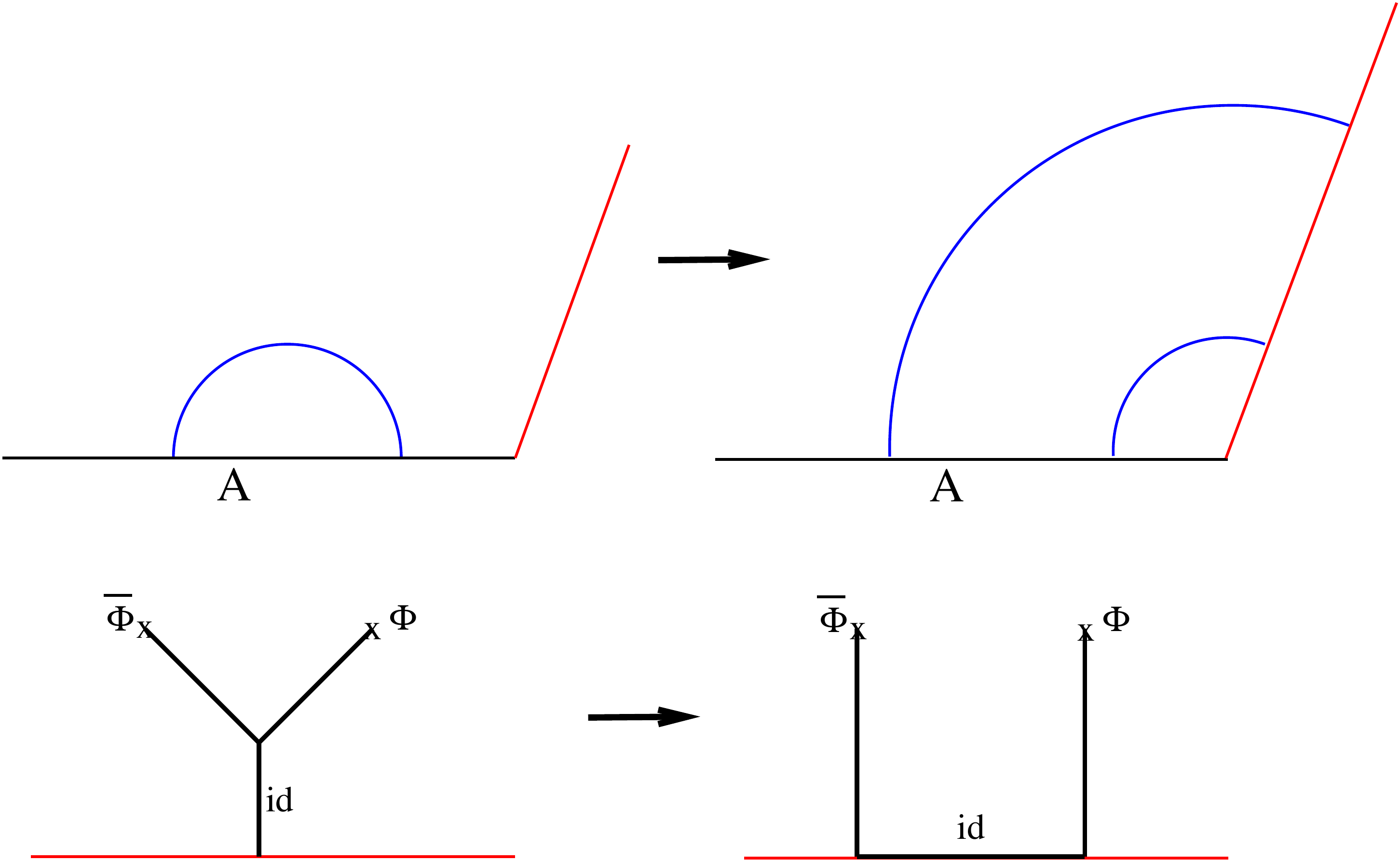}
  \caption{Top: transition in the RT surface from connected to disconnected topology; in black hole applications, the latter is associated with an entanglement wedge that includes the black hole interior. Bottom: BCFT interpretation in terms of the two-point function of twist operators used to compute entanglement entropy via the replica method. Phase transition comes from a switch of dominance between the identity block in a bulk channel to the identity block in a boundary channel.}
\label{fig:intro}
\end{figure}

The analysis in this paper parallels the computation of R\'{e}nyi entropies for multiple intervals in a CFT carried out by Hartman \cite{Hartman2013}. The CFT calculation makes use of the replica method, where we analytically continue the R\'{e}nyi entropies to find the entanglement entropy.
R\'{e}nyi entropies are computed from the CFT path integral on a multi-sheeted surface, or equivalently, from correlation functions of twist operators in replicated version of the original BCFT. 

These correlation functions can be expressed in terms of the basic BCFT data using operator product expansions.\footnote{Here, we have a bulk OPE, a bulk-boundary OE, and a boundary OPE.} Different ways of performing this expansion (different ``channels'') give the same result, but at large $c$, and under certain conditions on the CFT spectrum, the leading term in a $1/c$ expansion of the full result can be reproduced by truncating a single channel to include only intermediate states descending from the vacuum. Depending on the locations of the twist operators, this dominant channel can change, leading to a phase transitions in the entanglement entropy. The basic picture is illustrated in Figure \ref{fig:intro}.

A running theme is the use of the \emph{doubling trick}, and corresponding Ward identities, to simplify the kinematics of the BCFT. This relates the kinematics of BCFT $n$-point correlation functions on a half-plane to those of $2n$-point correlation functions in a chiral CFT on the full plane. This applies not only to the single interval case, where a two-point function in the BCFT is found to possess the same kinematics as a chiral four-point function, but to the multiple interval case, where the doubling trick can be applied to the monodromy method for computing conformal blocks.
%The extension of our results to multiple intervals relies on the monodromy method to analytically compute conformal blocks in the large $c$, single-replica limit.

\subsubsection*{Replica gravity calculation}

Faulkner \cite{Faulkner2013} simultaneously obtained the results in \cite{Hartman2013} from the gravity perspective.
Instead of calculating replica partition functions using large central charge CFT methods, the partition functions were calculated using gravitational path integrals in the dual Euclidean replica saddle geometries. We will see that Faulkner's calculation can be carried over directly to the calculation BCFT entanglement entropies. There, phase transitions in holographic entanglement entropy are mirrored by transitions in the dominant saddle contributing to the gravitational path integral. In the phase where the entanglement wedge of the radiation system includes the black hole interior, we can see explicitly from the Faulkner calculation that the relevant saddle is characterized by ``replica wormholes'', as suggested recently in  \cite{Almheiri2019qdq, Penington2019kki, Rozali2019}.

\subsubsection*{Outline}

In Section 2 below, we provide some basic background on BCFTs and recall various results that we will need for our calculation. In section 3, we recall the properties of gravity duals for holographic BCFTs and the holographic calculation of entanglement entropy in the various examples mentioned above. In Section 4, we present our main results,  the direct (B)CFT calculation of entanglement and Renyi entropies, and an analysis of the conditions on the spectum and OPE coefficients necessary to reproduce the gravity predictions. In Section 5, we generalize our BCFT results to the case of multiple intervals. Section 6 discusses the replica calculation of Renyi entropies on the gravity side and the relevance of replica wormholes. We finish with a discussion in Section 7.

%\section{Background on BCFTs}
\section{Review of boundary conformal field theory}
\label{sec:bcft}

In this section, we start with a brief review of boundary conformal field theories.\footnote{For more detailed reviews of BCFTs, see e.g. \cite{McAvity1995,  DiFrancesco1997, Cardy2004, Calabrese2009, Liendo2013,  Calabrese2016}.}

Given a CFT, we can define the theory on a manifold with boundary by making a choice of boundary conditions for the fields, and possibly adding boundary degrees of freedom coupled to the bulk CFT fields. For the theory defined on half of $\mathbb{R}^d$ or $\mathbb{R}^{d-1,1}$ (e.g. the region $x \ge 0$ for some spatial coordinate $x$), certain choices of the boundary physics give a theory that preserves the subset of the global conformal group mapping the half-space to itself, $\text{SO}(d,1) \subset \text{SO}(d+1,1)$ for the Euclidean case.\footnote{More generally, there can be boundary RG flows between such theories.} These choices define a \emph{boundary conformal field theory} (BCFT).\footnote{We can also consider the same theory with different boundary geometries; in this paper, we will only consider geometries that can be mapped to a half-space via a conformal transformation.} There are typically many choices of conformally invariant boundary condition for a given bulk CFT. We label the choice by an index $b$.

In this paper, we focus on BCFTs defined starting from  two-dimensional conformal field theories. In this case, there is a natural boundary analog of the central charge, known as the \emph{boundary entropy} $\log g_b$. This may be defined by considering the CFT on a half-space $x \ge 0$ with boundary condition $b$ at $x=0$. As we review in section 3.1 below, the entanglement entropy of an interval $[0,L]$ including the boundary is
\be
\label{Sint}
S = \frac{c}{6} \log {\frac{2 L}{\epsilon}} + \log g_b \; .
\ee
Thus, the boundary entropy gives a boundary contribution to the entanglement entropy. The quantity $g_b$ is also equal to the (regulated) partition function for the CFT on a disk with boundary condition $b$. %at the boundary.

\subsubsection*{Boundary states}

For any BCFT, there is a natural family of states $|b, \tau_0 \rangle$ that we can associate to the parent CFT defined on a unit circle. The wavefunctional $\langle \phi_0 |b, \tau_0 \rangle$ is defined as the Euclidean path integral for the CFT on a cylinder of height $\tau_0$, with boundary condition $b$ at Euclidean time $-\tau_0$ and CFT field configuration $\phi_0$ at $\tau = 0$.\footnote{With this definition, the norm of the states is not equal to 1.} We can formally define a \emph{boundary state} $|b \rangle$ associated with boundary condition $b$ via
\be
|b \rangle = |b, \tau_0 \to 0 \rangle \; .
\ee
In terms of $|b \rangle$, we have
\be
\label{boundarystate}
|b, \tau_0 \rangle = e^{- \tau_0 H} |b \rangle \; ,
\ee
since adding $\delta \tau$ to the height of the cylinder corresponds to acting on our state with Euclidean time evolution $e^{- \delta \tau H}$. The boundary state itself has infinite energy expectation value, but the Euclidean evolution used to define $|b, \tau_0 \rangle$ suppresses the high-energy components so that $|b, \tau_0 \rangle$ is a finite energy state.\footnote{This is a version of the \emph{global quench} considered in the condensed matter literature \cite{Calabrese2016}, but we have compactified the space on which the CFT is defined.} In general, this state is time-dependent.

The overlap of the boundary state $|b \rangle$ with the vacuum state is computed via the path integral on a semi-infinite cylinder. This can be mapped to the disk via a conformal transformation, so the result is the disk partition function:
\be
\langle 0 | b \rangle = g_b \; .
\ee

\subsubsection*{Boundary operators}

\label{sec:structure-constants}

In addition to the usual CFT bulk operators, a BCFT has a spectrum of local boundary operators $\hat{O}_J(x)$, each with a dimension $\hat \Delta_J$. Via the usual radial quantization (taking the origin to be a point on the boundary), these may be understood to be in one-to-one correspondence with the states of the BCFT on an interval with the chosen boundary condition at each end. The boundary operator dimension is equal to the energy of the corresponding state on the strip.

\subsubsection*{Symmetries and correlators}

A two-dimensional BCFT defined on the upper-half plane (UHP) preserves one copy of the Virasoro symmetry algebra, corresponding to transformations
\be
\delta z = \epsilon(z) \qquad \delta \bar{z} = \bar{\epsilon}(\bar{z}) \qquad  \bar{\epsilon}(\bar{z}) = \epsilon(\bar{z}^*) \; 
\ee
that map the boundary to itself.\footnote{Here, we recall that it is standard to treat $z$ and $\bar{z}$ as independent coordinates and consider a complexified version of the symmetry algebra for which the infinitesimal transformations are $\delta z = \epsilon(z)$ and $\delta \bar{z} = \bar{\epsilon}(\bar{z})$. The non-complexified transformations correspond to taking $\bar{\epsilon}(\bar{z}) = \epsilon(\bar{z}^*)$ with $\epsilon(x)$ real for real $x$, or $\bar{\epsilon}(\bar{z}) = -\epsilon(\bar{z}^*)$ with $\epsilon(x)$ pure imaginary for real $x$. Of these, the first set preserves the upper half plane, acting explicitly as $\delta x = (\epsilon(x+iy) + \epsilon(x-iy))/2, \delta y = -i(\epsilon(x+iy) - \epsilon(x-iy))/2$ on the physical coordinates.} These correspond to a set of generators
\be
\tilde{L}_n = L_n + \bar{L}_n \; .
\ee

In this case, the conformal Ward identity becomes
\beas
    &&\langle \tilde T(z) \prod_i {\cal O}_{h_i \bar h_i}(z_i,\bar{z}_i) \rangle  \cr
    && \qquad \qquad = \sum_i \left( \frac{h_i}{(z-z_i)^2} + \frac{1}{z-z_i}\frac{\partial}{\partial z_i} + \frac{\bar h_i}{(\bar z- \bar z_i)^2} + \frac{1}{\bar z-\bar z_i}\frac{\partial}{\partial \bar z_i} \right) \langle \prod_i {\cal O}_{h_i \bar h_i}(z_i,\bar{z}_i) \rangle \, ,
\eeas
where $\tilde T(z) = \sum_n z^{-n-2} \tilde{L}_n$.

The Virasoro symmetry algebra of a BCFT is thus the same as that of a chiral CFT on the whole plane. A consequence is that the kinematics (i.e. the functional form of correlators given the operator dimensions) of the BCFT in the UHP is directly related to that of a chiral CFT on the whole plane. Correlators
\be
\label{UHP}
\uavg{{\cal O}_{h_1 \bar h_1}(z_1,\bar{z}_1) \cdots {\cal O}_{h_n \bar h_n}(z_n,\bar{z}_n) } 
\ee
of bulk CFT operators ${\cal O}_{h_k \bar h_k}$ with conformal weights $(h_k,\bar{h}_k)$ in the UHP are constrained to have the same functional form as chiral CFT correlators 
\be
\label{ChiCorr}
\langle {\cal O}_{h_1}(z_1) \cdots {\cal O}_{h_n}(z_n) {{\cal O}}_{\bar h_1}(\bar{z}_1) \cdots {{\cal O}}_{\bar h_n}(\bar{z}_n) \rangle
\ee
of fields ${\cal O}_{h_k}$ and ${{\cal O}}_{\bar h_k}$ with chiral weights $h_k$ and $\bar{h}_k$ respectively.\footnote{Here, the original theory is defined on the slice where $\bar z = z^*$, so the operators ${{\cal O}}_{\bar h_i}(\bar{z}_i)$ live on the lower half-plane.} 
More generally, we can include boundary operators $\hat{ \cal O}_{\hat \Delta_I}(x_I)$ in (\ref{UHP}), where $x_I$ is real. In this case, the functional form is reproduced by adding chiral operators with $h_I = \hat \Delta_{I}$ at $z = x_I$ to the chiral correlator (\ref{ChiCorr}).  See \cite{Recknagel:2013uja} for a more complete discussion of this constraint, often referred to as the ``doubling trick''. 

We will later make use of this kinematic equivalence to relate conformal blocks for a BCFT on the UHP to chiral conformal blocks on the entire plane.

\subsubsection*{Bulk one-point functions}

The doubling trick implies that a primary operator with weights $(h,\bar{h})$ is kinematically allowed to have a nonvanishing one-point function if $h= \bar{h}$ (i.e. for a scalar primary). In this case, the one-point function $\uavg{\Ocal_{h,h}(z, \bar{z})}$ is constrained to have the same form as a chiral two-point function $\langle {\cal O}_h(z)  \bar{{\cal O}}_h(z^*) \rangle$, so we have
\begin{equation}
  \label{1-pt-uhp}
  \uavg{ \Ocal_{h, \bar{h}}(z, \bar{z})} =
  \frac{\mathcal{A}^b_\Ocal}{|z-z^*|^{2h}} =\frac{\mathcal{A}^b_\Ocal}{|2y|^{\Delta_\Ocal}}.
\end{equation}
where we take $z = x + i y$ here and below.
Once the normalization of the operators is fixed by choosing the normalization of the two-point function in the parent CFT, the coefficient $\mathcal{A}^b_\Ocal$ in the one-point function is a physical parameter that depends in general on both the operator and the boundary condition. 

Here and everywhere in this paper we will take the expectation value $\uavg{\cdot}$ to be normalized by the UHP partition function so that
\begin{equation}
    \uavg{\mathbf{1}} = 1 \, .
\end{equation}

\subsubsection*{Bulk-boundary two-point functions}

The correlation function 
\be
\uavg{\Ocal_i(z, \bar{z}) \hat{ \cal O}_I(x')}
\ee
of bulk and boundary primary operators is constrained to have the functional form of a chiral three-point function
\be
\langle {\cal O}_{h_i}(z) {\cal O}_I(x') {\cal O}_{\bar{h}_i}(\bar{z}) \rangle \; .
\ee
For a scalar operator ${\cal O}_i$, this gives
\be
\label{boundarybulk}
\uavg{ {\cal O}_i (z,\bar{z}) \hat{ \cal O}_I(x')} = \frac{\Bcal^b_{iI}}{(2 y)^{\Delta_i - \Delta_I}(y^2 + (x-x')^2)^{\Delta_I}}\;,
\ee
where $\Bcal^b_{iI}$ forms part of the basic data of our BCFT. Taking $\hat {\cal O}_I$ to be the identity operator, we have  from the previous section that $\Bcal^b_{i 1} = \mathcal{A}^b_i$.

\subsubsection*{Boundary operator expansion and OPEs}

In the same way that a pair of bulk operators at separated points can be expanded as a series of local operators via the OPE, a bulk operator can be expanded in terms of boundary operators via a \emph{boundary operator expansion (BOE)}.\footnote{This follows by the same logic of the state-operator mapping and OPE in a CFT. The state produced by a bulk operator can be mapped by an infinite dilation to a local operator at the origin on the boundary. And, as in the OPE, we choose to expand this local operator in terms of a basis of dilation eigenstates. }
For a scalar primary operator, symmetries constrain the general form of this expansion to be
\bea
    \Ocal_i(z, \bar{z}) &=& \sum_{J} {\Bcal^{b J}_{i}  \over (2y)^{\Delta_i-\Delta_I}}
    \tilde{C}[y,\partial_{x}]\hat{\mathcal{O}}_{J}(x) \cr
   &=&  \sum_{J} {\Bcal^{b J}_{i}  \over (2y)^{\Delta_i-\Delta_I}}
    \hat{\mathcal{O}}_{J}(x) + {\rm desc.}\;,
    \label{boe-2}
\eea
where the sum is over boundary primary operators. The differential operator $\tilde{C}$ determines the contribution of descendant operators and depends only on the conformal weights of $\Ocal_i$ and $\hat{\mathcal{O}}_{J}$. The coefficients $\Bcal^{b J}_{i}$ are related to the ones appearing in the bulk-boundary two-point function by raising the index with the metric $g_{IJ}$ appearing in the boundary two-point function 
\be
\langle \hat \Ocal_I (x_I) \hat \Ocal_J (x_I) \rangle = {g_{IJ} \over |x_I - x_J|^{2 \Delta_I}} \; ,
\ee
though we will generally assume that we are working with a basis of boundary operators for which $g_{IJ} = \delta_{IJ}$.

Below, we will also make use of the ordinary OPE for bulk scalar operators,\footnote{To see that this should still be valid in the presence of a boundary, note that in a conformal frame where the upper-half-plane is mapped to the exterior of a circle surrounding the origin, the presence of the boundary is equivalent to the insertion of an operator at the origin (specifically, the operator associated with the state $|b, \tau_0 \rangle$ described above).}
\bea
    \label{ope}
    \Ocal_i(z_1, \bar{z}_1) \Ocal_j(z_2, \bar{z}_2) &=& \sum_k {\hat{\Ccal}_{ij}^{k} \over |z_1 - z_2|^{\Delta_i + \Delta_j - \Delta_k}}  C_{\Delta_i \Delta_j ; \Delta_k}[z_{12},\partial_{z}] \Ocal_k(z_2, \bar{z}_2) \cr
    &=& \sum_k {\hat{\Ccal}_{ij}^{k} \over |z_1 - z_2|^{\Delta_i + \Delta_j - \Delta_k }}  \Ocal_k(z_2, \bar{z}_2) + {\rm desc.}
\eea
Finally, there is also an OPE for boundary fields, but we will not need this in our calculations below.

\subsubsection*{Two-point functions and conformal blocks}

We now consider the bulk two-point function. Here, we restrict to scalar primary operators of equal dimension $\Delta$ since that is what we will need below. However, in general, bulk two-point functions in a BCFT can be non-vanishing for any conformal weights $(h_1,\bar{h}_1)$ and $(h_2,\bar{h}_2)$. We discuss the general case in detail in Appendix A. 

By the doubling trick, the BCFT two-point function
\begin{equation}
   \uavg{ \Ocal_1 (z_1, \bar{z}_1) \Ocal_2 (z_2, \bar{z}_2)} \, 
\end{equation}
of scalar operators with dimension $\Delta$ has the same functional form as a four-point function of chiral operators 
\begin{equation}
   \langle \Ocal_1(z_1)  \Ocal_2(z_2) \Ocal_3(\bar{z}_2) \Ocal_4 (\bar{z}_1) \rangle \, ,
\end{equation}
where each operator has chiral weight $h = \Delta/2$. Making use of (\ref{chiral}) for the general form of such a correlator, we have that
\be
\label{BCFT2pt}
  \uavg{ \Ocal_1 (z_1, \bar{z}_1) \Ocal_2(z_2, \bar{z}_2)} = \left[ {\eta \over 4 y_1 y_2 }\right]^{\Delta} F(\eta)
\ee
where $F(\eta)$ is some function of the cross-ratio 
\be
\eta = {(z_1 - \bar{z}_1)(z_2- \bar{z}_2) \over (z_1 - \bar{z}_2)(z_2 - \bar{z}_1)} \; .
\ee
The function $F$ can be written more explicitly by making use of either the BOE %boundary operator expansion 
or the bulk OPE for the operators in  (\ref{BCFT2pt}). 
Using the BOE %boundary operator expansion 
for each operator in (\ref{BCFT2pt}),  the bulk two-point functions can be expressed as a sum of boundary two-point functions. In this way, the function $F(\eta)$ in (\ref{BCFT2pt}) may be expressed as
\be
\label{2ptBoundary}
F(\eta) = \sum_I \mathcal{B}^{b I}_{{\cal O}_1} \mathcal{B}^b_{{\cal O}_2 I} \mathcal{F}(c,\Delta_I ,\Delta/2%{\Delta \over 2}
\,|\, \eta)
\ee
where the sum is over boundary primary operators and $\mathcal{F}(c,\Delta_I ,\Delta/2 %{\Delta \over 2}
| \eta)$ gives the contribution of a single boundary primary operator and all of its Virasoro descendants. We show in Appendix A that this function is %exactly
the usual conformal block appearing in the expansion of a chiral four-point function of operators with equal conformal weight $h = \Delta/2$.\footnote{In general, the conformal block depends on four external weights; here and below, we will use the shorthand $\mathcal{F}(c,h_{\text{int}}, h| \eta) \equiv \mathcal{F}(c,h_{\text{int}}, [h,h,h,h]| \eta)$ where the latter is the general expression for the chiral conformal block used in Appendix A.} 

We can alternatively use the bulk OPE to reduce the BCFT two-point function (\ref{BCFT2pt}) to a sum of one-point functions. This leads to an alternative expression for $F(\eta)$, 
\be
\label{2ptBulk}
F(\eta) = \sum_i \Ccal^i_{{\cal O}_1 {\cal O}_2} \Acal^b_i \mathcal{F}(c,\Delta_I ,\Delta/2\,|\,1 - \eta)
\ee
Here, $\mathcal{F}$ is the same chiral conformal block as in (\ref{2ptBoundary}), as we show in Appendix A.
The equivalence of the expressions (\ref{2ptBoundary}) and (\ref{2ptBulk}) is a BCFT version of the usual crossing symmetry constraints; in this case, we have a relation between bulk OPE coefficients and boundary operator expansion coefficients.

\section{Holographic BCFT entanglement entropies}

In this section, we review the holographic calculation of entanglement entropies for subsystems of BCFTs with gravitational duals, or for states of holographic CFTs defined via Euclidean BCFT path integrals. These are the results that we will try to understand via direct CFT calculations in the next section.

\subsection{Holographic BCFTs}

Certain BCFTs have a dual gravitational description. These correspond to holographic CFTs defined on a space $M$ with boundary $\partial M$, and a boundary condition perhaps obeying  additional constraints so that the theory remains holographic. %with some choice of boundary condition, perhaps itself constrained in some way to be holographic. 
The dual geometries are asymptotically AdS with boundary geometry $M$,
%, such that the boundary geometry is $M$. 
but the bulk physics associated with $\partial M$ can be different depending on the choice of boundary condition.

For a $d$-dimensional CFT, we can have an effective ``bottom up" description of the gravity dual as a $(d+1)$-dimensional asypmtotically AdS spacetime with an end-of-the-world (ETW) brane extending from $\partial M$
%out from the boundary of the space on which the CFT lives
\cite{Karch:2000gx,Takayanagi:2011zk,Fujita:2011fp,Astaneh:2017ghi}. 
However, in  ``top down" microscopic examples (see for instance \cite{Chiodaroli:2011fn, Chiodaroli:2012vc, DHoker:2007zhm, DHoker:2007hhe, Aharony:2011yc,Assel:2011xz}), the dual can be a smooth higher-dimensional geometry. In this case, the ETW brane in the lower-dimensional description represents the smooth degeneration of an internal dimension. %. The ETW brane in the lower-dimensional effective description represents a region of the higher-dimensional geometry where an internal space smoothly degenerates.

The simplest possible gravitational dual has
%dual gravitational description is %an effective theory with an
an ETW brane coupling only to the bulk metric field. Its action is taken to include a boundary cosmological constant (interpreted as the brane tension) and a Gibbons-Hawking term involving the trace of the extrinsic curvature. The details of the action and equation of motion, and all the solutions that we will require in this paper, may be found in \cite{Cooper2018}.
A more general ansatz is an ETW brane action with coupling to additional bulk fields, e.g. light scalars. %More generally, we can consider ETW brane actions with couplings to additional bulk fields.

\subsection{Entanglement entropies for holographic BCFTs}

We can use the Ryu-Takayanagi (RT) formula \cite{Ryu2006a} to holographically calculate the entanglement entropy for spatial subsystems. As usual, the entropy (at leading order in the $1/c$ expansion) is given as
\be
S_A = {1 \over 4 G} {\rm Area}(\tilde{A})\;,
\ee
where $\tilde{A}$ is the minimal area extremal surface in the dual geometry homologous to the boundary region $A$.

%In computing the entanglement entropy of a subsystem of a holographic BCFT, a new feature is that we can have RT surfaces that end on the ETW brane. 
A new feature of entanglement entropy for holographic BCFTs is that the RT surfaces can end on the ETW brane \cite{Takayanagi2011}. 
Here, we should keep in mind that the ETW brane itself represents a part of the bulk geometry. The homology condition says that the RT surface $\mathcal{X}_A$ for a region $A$ on the boundary, together with the region $A$ itself, should be the boundary of a region $\Xi_A$ of the bulk spacetime: $\partial \Xi_A = A \cup \mathcal{X}_A$. But when applying this condition, the ETW brane should be considered as part of this bulk spacetime region $\Xi_A$, rather than an additional contribution to the boundary. As a result, we can have a disconnected RT surface for a connected boundary region, as shown in figure 3.

\subsubsection{BCFT vacuum state on a half space}
\label{sec:half-space}

As an example, consider the vacuum state of a two-dimensional BCFT on a half-space $x > 0$. Here, the $\text{SO}(1,2)$ symmetry preserved by the BCFT should be reflected in the dual geometry. Generally, this gives a warped product of AdS$_2$ and an internal space, such that the full geometry has an asymptotic region that is locally AdS$_3$ times some internal space, with boundary geometry equal to the half-space on which the CFT lives. In general, we can write the metric as
\begin{equation}
\label{metric-bads-1}
  \D s_{\mathcal{M}}^2 = {\ell_\AdS^2}\left[\hat{g}_{ij}(\mu) \D\mu_i \D\mu_j + \frac{f(\mu)}{z ^2} (\D
  z^2-\D t^2)\right].
\end{equation}
Microscopic solutions of this type were constructed in \cite{Chiodaroli:2011fn, Chiodaroli:2012vc}.

We can also give a lower dimensional description (at least in the vicinity of the boundary), where we reduce on the internal space so that the internal metric is represented via scalars and vectors. In this case, we can write
\begin{equation}
\label{metric-bads}
  \D s_{\mathcal{M}}^2 = {\ell_\AdS^2}\left[\D\mu^2 + \frac{f(\mu)}{z ^2} (\D
  z^2-\D t^2)\right],
\end{equation}
where $f(\mu) \to \cosh^2(\mu/\ell_\AdS)$ as we approach the asymptotic boundary at $\mu = - \infty$ so that the metric is asymptotically AdS$_3$. In general, the scalar fields in the geometry can be functions of the coordinate $\mu$.

\begin{figure}[t]
  \centering
  \includegraphics[scale=0.2]{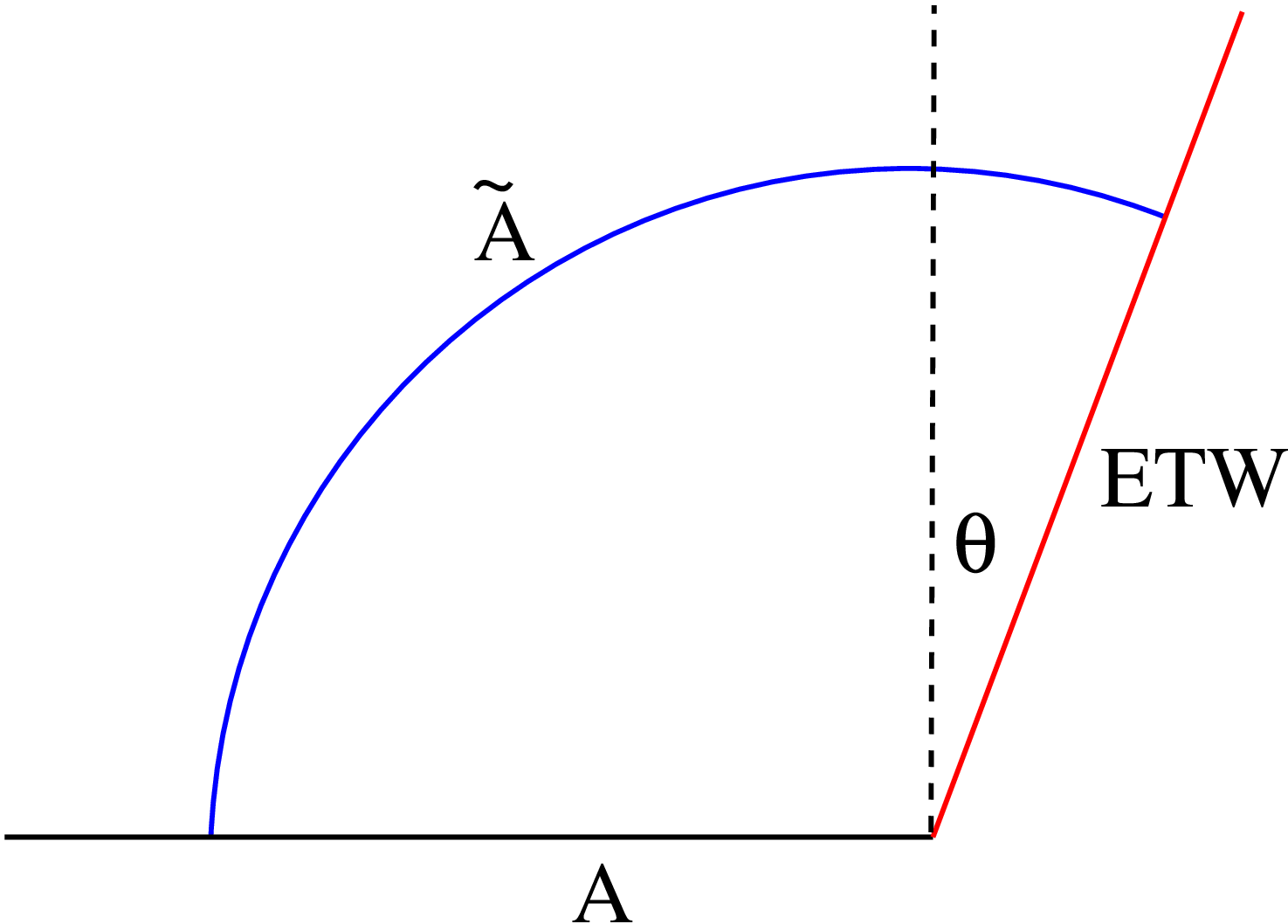}
  \caption{Holographic calculation of entanglement entropy for an interval $A$ containing the boundary. The RT surface $\tilde{A}$ sits at a fixed location on the AdS$_2$ fibers of the dual geometry. Here $A$ is homologous to $\tilde{A}$ since the ETW brane represents a smooth part of full microscopic geometry.}
\label{fig:RT1}
\end{figure}

In the simplest effective bulk theory, there is an ETW brane with stress-energy tensor $8 \pi G T_{ab} = -T g_{ab}/\ell_\AdS$ \cite{Karch:2000gx,Takayanagi:2011zk}, and bulk geometry pure AdS, with $f(\mu) = \cosh^2(\mu/\ell_\AdS)$. The brane sits at $\mu_{\text{max}} = {\rm arctanh}(T)$. %\footnote{
Here, the coordinate $\mu$ is related to the angular coordinate $\theta$ in a polar-coordinate description of Poincar\'e-AdS by $1/\cos(\theta) = \cosh(\mu)$, so the brane goes into the bulk at a constant angle $\theta = \arcsin(T)$, as shown in Figure \ref{fig:RT1}.
%}

\subsubsection*{Entanglement entropy for an interval including the boundary}

We now consider the entanglement entropy for an interval in the half-space. In the case of an interval $[0,L]$ containing the boundary, we expect the universal form (\ref{Sint}) for the entanglement entropy. In the holographic calculation with the general metric (\ref{metric-bads-1}), the RT surface sits at a constant position on the AdS$_2$ fiber, so the entanglement entropy is
\be
S = {1 \over 4 G} \int_{z > \epsilon} \D^{d+1} x \sqrt{\hat{g}}\;,
\ee
where $z$ is the Fefferman-Graham radial coordinate and $d$ is the dimension of the internal space. We can regulate this by subtracting off half the area of the entangling surface of an interval of length $2L$ in vacuum AdS, so
\be
S = \left[ {\ell^{d+1}_\AdS \over 4 G} \int_{z > \epsilon} \D^d \mu_i \sqrt{\hat{g}} - {\ell^{d+1}_\AdS \over 4 G} \int_{z > \epsilon, x > 0} \D^d x \sqrt{\hat{g}_\AdS}\right] + {\ell_\AdS^{d+1} \over 4 G} \int_{z > \epsilon, x > 0} \D^d x \sqrt{\hat{g}_\AdS}\;,
\ee
where $\hat{g}_\AdS$ is the metric for pure $AdS$. In this expression, the term in square brackets has a finite limit as $\epsilon \to 0$, independent of $L$, while the second term gives ${c \over 6} \log \left( {2 L \over \epsilon} \right)$. Thus, we reproduce (\ref{Sint}), with the identification
\be
\log g_b = \lim_{\epsilon \to 0}  \left[ {\ell_{\AdS}^{d+1} \over 4 G} \int_{z > \epsilon} \D^d x \sqrt{\hat{g}} - {\ell_{\AdS}^{d+1} \over 4 G} \int_{z > \epsilon, x > 0} \D^d x \sqrt{\hat{g}_\AdS}\right] \; .
\ee
As an example, with a constant tension ETW brane, we have
\be
\log g_b = {\ell_\AdS \over 4 G} \int_0^{\mu_{\text{max}}} \D \mu = {\ell_\AdS \over 4 G} {\rm arctanh}(T) = {c \over 6} {\rm arctanh}(T) \; .
\ee
This is the result of Takayanagi \cite{Takayanagi2011} relating the boundary entropy to brane tension.

\subsubsection*{Entanglement entropy for an interval away from the boundary}

\begin{figure}[t]
  \centering
  \includegraphics[scale=0.2]{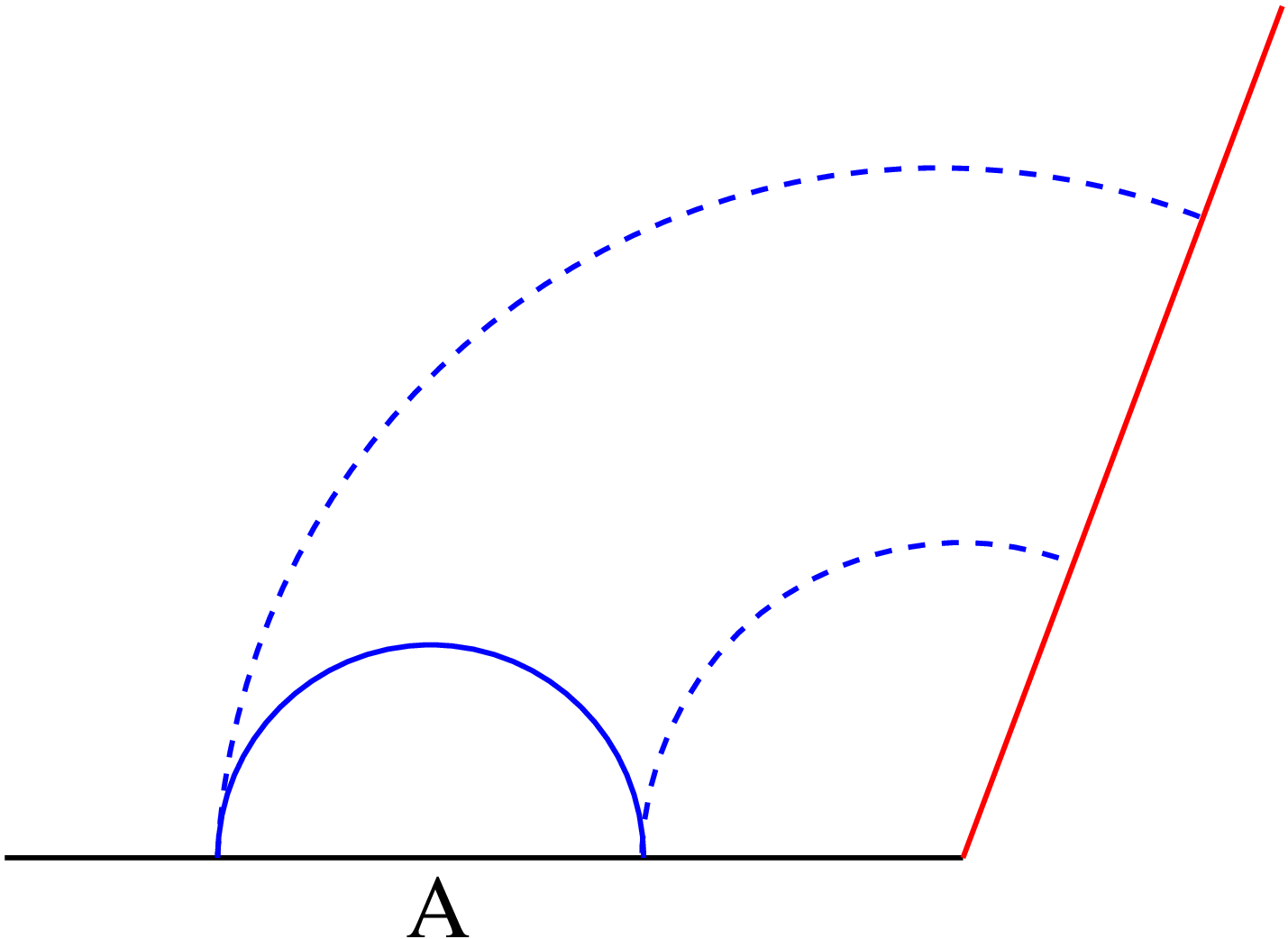}
  \caption{Holographic calculation of entanglement entropy for an interval $A$ away from the boundary. The RT surface has two possible topologies, a connected  (solid curve) and  disconnected (dashed curves).}
\label{fig:RT12}
\end{figure}

Now consider the holographic calculation of entanglement entropy for an interval $[x_1,x_2]$ away from the boundary. 
In general, the CFT result for this entanglement entropy does not have a universal form.

In the holographic calculation, we can have a phase transition between two different possible RT surface topologies: a connected RT surface or a disconnected surface with both components ending on the ETW brane. The two topologies are shown in Figure \ref{fig:RT12}.

Let us consider these phases in more detail.
In the disconnected case, the RT surface computing the entanglement entropy of an interval $[x_1,x_2]$ is the union of the RT surfaces associated with $[0,x_1]$ and $[0,x_2]$. % each including the boundary.
Thus, to leading order in large $N$, we have that
\be
\label{disc}
S_{[x_1,x_2]}^{\text{disc}} = S_{[0,x_1]} + S_{[0,x_2]} = {c \over 6} \log \left( {2 x_1 \over \epsilon} \right) + {c \over 6} \log \left( {2 x_2 \over \epsilon} \right) + 2 \log g_b \; .
\ee
This result makes use only of the disconnected topology of the RT surface, so is a universal result for the disconnected phase in any holographic theory.

In the connected phase (expected to apply when the interval is sufficiently far from the CFT boundary), there is in general no simple universal result for the entanglement entropy. We need to find an RT surface in the dual geometry, and the calculation of this surface will depend on the details of the metric $\hat{g}$ appearing in (\ref{metric-bads-1}).

For certain boundary conditions, it may be that the dual gravitational theory is well-described by an ETW brane with only gravitational couplings. In this case, the dual geometry is locally AdS$_3$, and the calculation of entanglement entropy for the interval will be the same as the holographic calculation of vacuum entanglement entropy for the same interval in the CFT without a boundary. Thus, we have
\be
\label{conn}
S_{[x_1,x_2]}^{\text{conn}} = {c \over 3} \log \left({x_2 - x_1 \over \epsilon}\right)\;. %\qquad \qquad {\rm gravitational \; ETW \; brane} \; .
\ee
Below, we will try to understand what conditions must be satisfied in the BCFT in order that this result is correct.

In cases where (\ref{disc}) and (\ref{conn}) give the correct results for the two possible RT-surface topologies, the actual entanglement entropy will be computed by taking the minimum of these two results. We find that the disconnected surface gives the correct result for the entanglement entropy when
\be
\label{transition}
\log \left[{1 \over 2} \left(\sqrt{x_2 \over x_1} - \sqrt{x_1 \over x_2} \right) \right] >  {6 \log g_b \over c}\;,
\ee
so that for a fixed interval size, we have a phase transition as the location of the interval relative to the boundary is varied.

In the more general case where the bulk geometry is not locally AdS, there is no explicit result for the entanglement entropy in the connected phase and (\ref{transition}) does not apply. However, we expect that the qualitative behavior of the entanglement entropy is similar, with a transition to the disconnected phase as the interval approaches the boundary. We can view this as a prediction for the behavior of entanglement entropy in holographic BCFTs. One of our main goals below will be to understand the existence of this transition via a direct CFT calculation.

Before turning to the machinery of CFTs, we review two closely related holographic calculations. In both cases, the dual geometry involves a black hole, and the transition in RT surfaces takes us between phases where the entanglement wedge of the CFT region under consideration does or does not include a portion of the black hole interior.

\subsubsection{Entanglement entropy for boundary states $|b, \tau_0 \rangle$}
\label{sec:quench}

%In our second example, we consider the entanglement entropy of an interval of angular size $\Delta \theta$ for a CFT on $S^1$ in a state $|b, \tau_0 \rangle$ defined in (\ref{boundarystate}) via the Euclidean path-integral on a finite cylinder with boundary condition $b$ applied at past Euclidean time $-\tau_0$.

\begin{figure}
  \centering
  \includegraphics[scale=0.2]{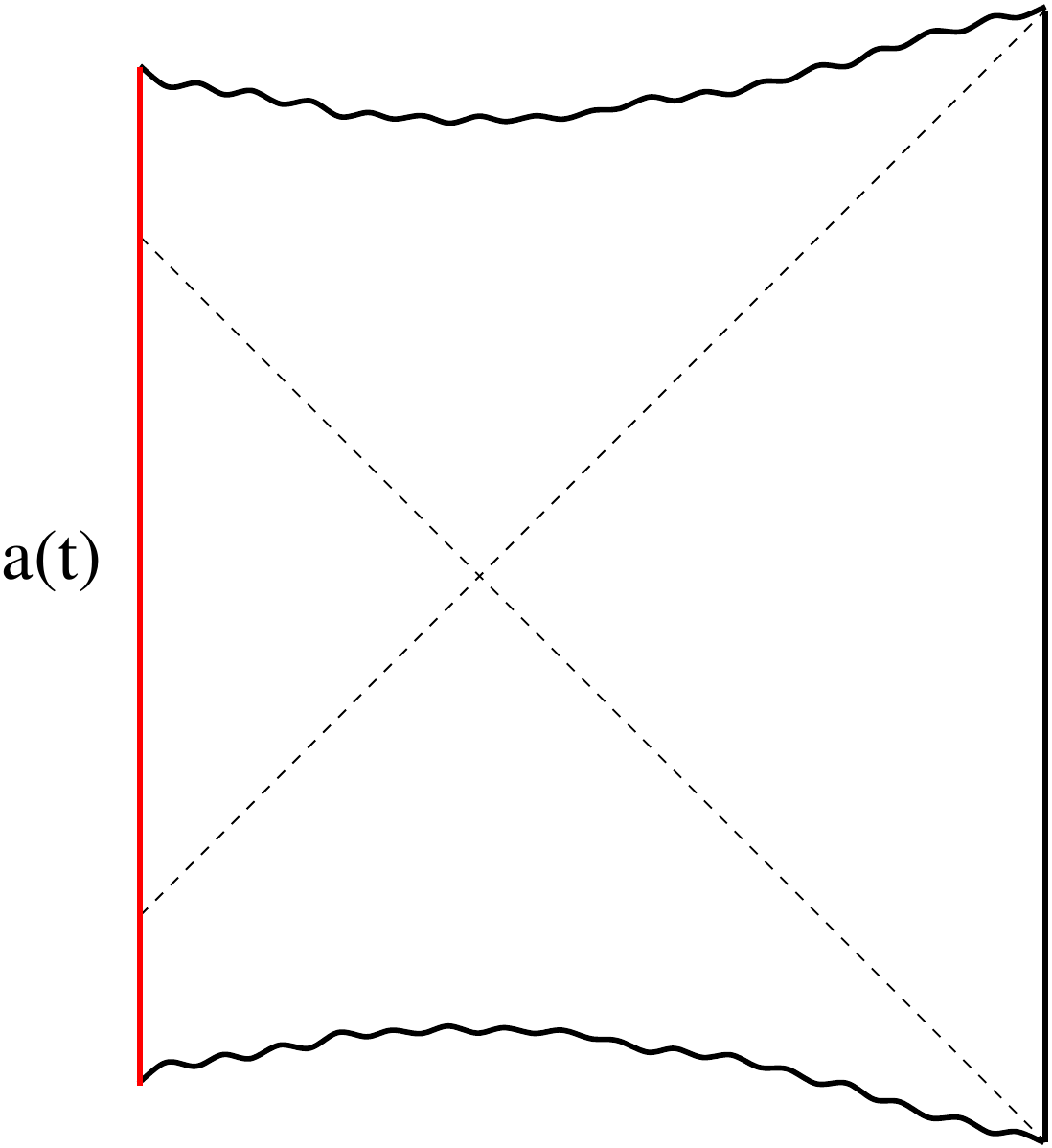}
  \hskip 0.5 in
  \includegraphics[scale=0.2]{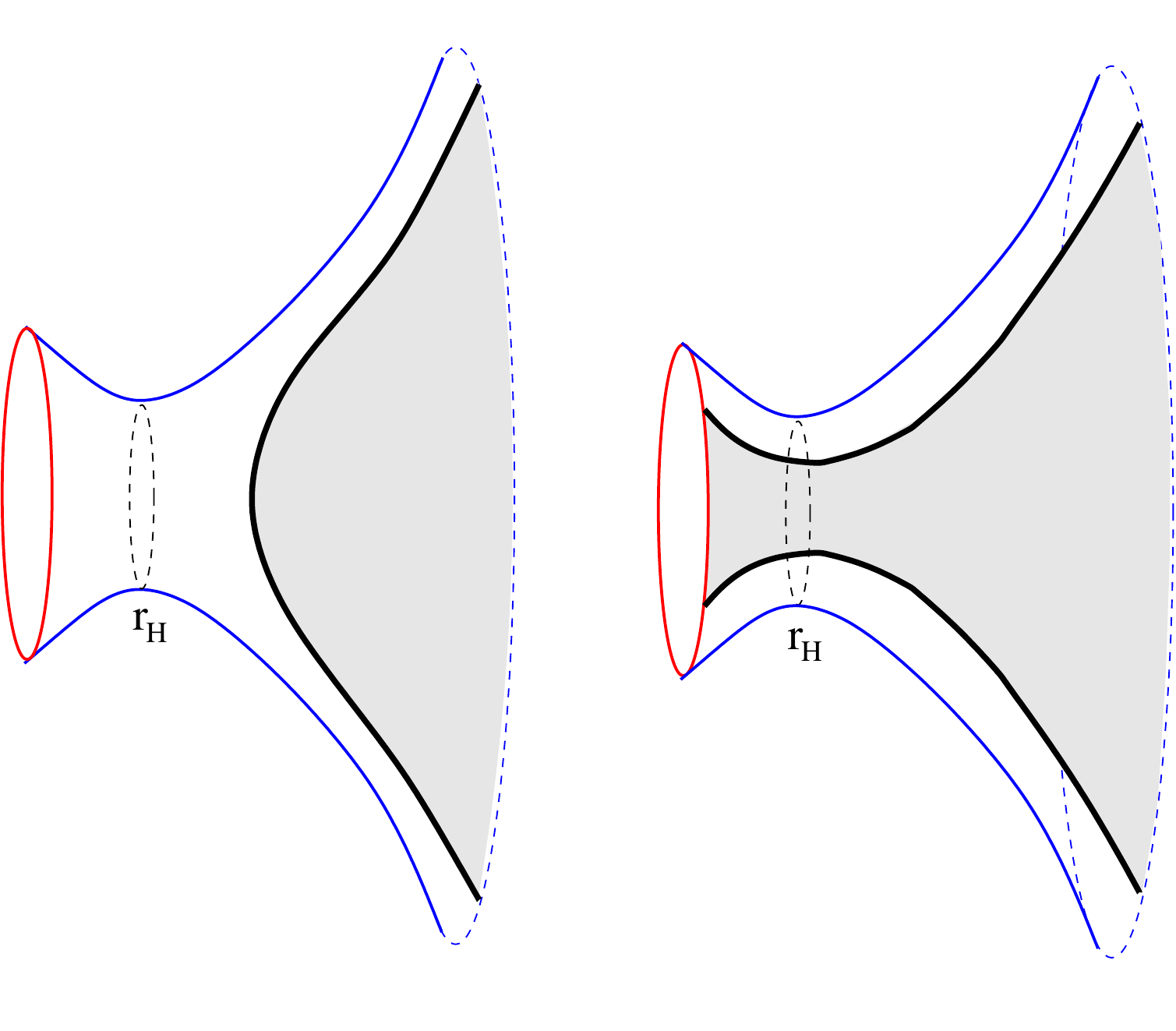}
  \caption{The dual geometry for $|b,\tau_0\rangle$ for sufficiently small $\tau_0$ is a portion of the maximally-extended AdS-Schwarzchild geometry, cut off by a spherically symmetric ETW brane. The pictures on the right show the spatial slice at $t=0$ and the connected and disconnected topologies for the RT surface corresponding to a large interval on the boundary circle.}
\label{fig:microstate}
\end{figure}

Consider a CFT on $S^1$, in the state $|b, \tau_0 \rangle$ defined via the Euclidean path integral (\ref{boundarystate}). We can consider the entanglement entropy for an interval of angular size $\Delta \theta$ in this state, at some fixed Lorentzian time.
As described in \cite{Almheiri2018,Cooper2018}, for small enough $\tau_0$, this is a high-energy pure state of the CFT and the dual geometry is expected to be black hole. Assuming that the bulk effective gravitational theory for the BCFT involves a purely gravitational ETW brane of tension $T$, %such that the bulk effective gravitational theory involves a purely gravitational ETW brane with tension $T$, 
it was shown in \cite{Almheiri2018,Cooper2018} that the dual geometry for $T > 0$ is a portion of the maximally-extended AdS-Schwarzchild geometry. The black hole interior terminates on a spherically-symetric ETW brane with a time-dependent radius, as shown in Figure \ref{fig:microstate}.

In this case, the geometry outside the horizon is pure AdS-Schwarzchild. In the connected phase, which dominates for small enough $\Delta \theta$, the RT surface lies entirely outside the horizon and gives a time-independent entanglement entropy
\be
\label{BSconn}
S^{\text{conn}} = {c \over 3} \log \left[ {4 \tau_0 \over \pi \epsilon} \sinh \left( \pi \Delta \theta \over 4 \tau_0 \right) \right]\;,
\ee
where we take the circumference of the CFT circle to be 1. 

For small enough $\tau_0$, large-enough interval size $\Delta\theta$, and time $t$ sufficiently close to 0 (when the state is prepared) we also have a disconnected phase, where the RT surface is a union of two surfaces at fixed angular position that enter the horizon and terminate on the ETW brane. Here, we find that 
\be
\label{BSdisc}
S^{\text{disc}} = {c \over 3} \log \left[ {4 \tau_0 \over \epsilon \pi } \cosh \left({\pi t \over 2 \tau_0} \right)\right] + 2 \log g_b \; .
\ee
This is smaller than the connected result (and thus represents the actual entanglement entropy) when
\be
\sinh \left( \pi \Delta \theta \over 4 \tau_0 \right) \ge \cosh \left({\pi t \over 2 \tau_0} \right) e^{6 \log g_b \over c} \; .
\ee
When this condition is satisfied, the entanglement wedge of the interval includes a portion of the black hole interior, and hence the entanglement entropy probes the interior geometry. For late times, the connected phase always dominates. This is consistent with the expectation that the state will thermalize, so that the entanglement entropy for a subsystem gives the thermal result.\footnote{This is similar to the behaviour of correlators in microscopic models of black hole collapse  based on approximate global quenches of Vaidya type \cite{Anous2016}. A phase transition in channel dominance leads to a shift in the gravitational saddle computing entanglement entropy, which in turn is responsible for maintaining unitarity. We thank Tarek Anous for discussion of this point.} % after enough time has passed.

\subsubsection{Entanglement entropy in the thermofield double state of two BCFTs}
\label{sec:tfd}

In our final example, we take the thermofield double state of two BCFTs, each on a half-space, and consider the entanglement entropy for the subsystem $A(x_0)$ consisting of the union of the regions $[x_0, \infty)$ in each CFT, as in Figure \ref{fig:CFTpics}d.
In \cite{Rozali2019}, following \cite{Almheiri2019a}, it was argued that this system provides a model of a two-sided 2D black hole coupled to an auxiliary radiation system, where the Page time for the black hole is $t_{\text{Page}} \sim 6\log(g_b)/c$. 

While the simple observables in this system are time-independent, the holographic calculations in \cite{Rozali2019} revealed that the entanglement entropy for the subsystem $A(x_0)$ increases with time, then undergoes a phase transition. After this transition, the entanglement entropy is time-independent and the entanglement wedge of the radiation system includes a substantial portion of the black hole interior. The interpretation is that while no net energy is exchanged between the black hole and the radiation system, information from the black hole escapes into the radiation system until the radiation system contains enough of it % information 
to reconstruct the black hole interior.

\begin{figure}
  \centering
  \includegraphics[scale=0.3]{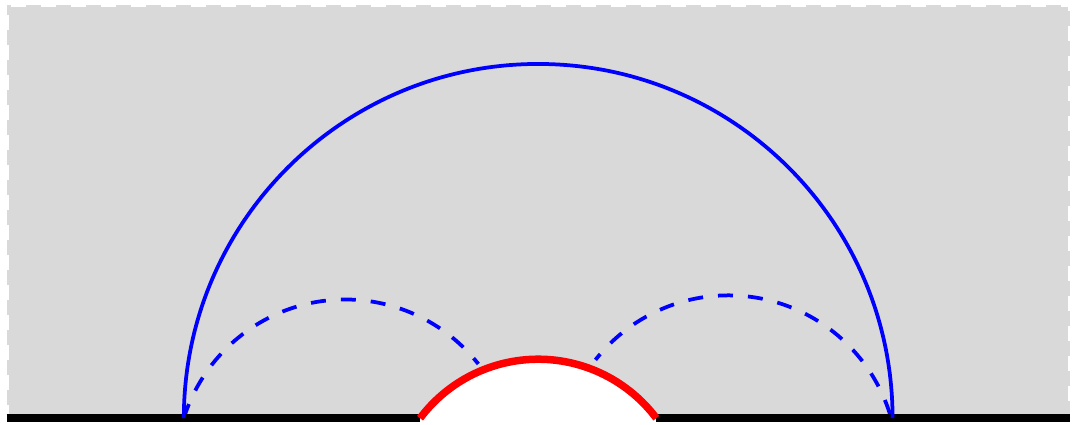}
  \caption{Dual geometry to the thermofield double state of two BCFTs, showing two possible topologies for the RT surface for the region corresponding to the union of points in either CFT at a distance larger than $x_0$ from the boundary.}
\label{fig:TDphases}
\end{figure}

The phase transition is seen most easily via a holographic calculation in the dual three-dimensional gravity picture. Here, we have a dynamical ETW brane that connects the two CFT boundaries, as shown in Figure \ref{fig:TDphases}. At early times, the RT surface for $A(x_0)$ is connected and does not intersect the ETW brane. The entanglement entropy is
\be
\label{BRconn}
S^{\text{conn}} = {c \over 3} \log \left({2 \over \epsilon} \cosh t \right) \; .
\ee
At late times, the RT surface is disconnected with components stretching from the boundary of each BCFT to the ETW brane. In this case, the entropy is 
\be
\label{BRdisc}
S^{\text{disc}} = {c \over 3} \log \left({2 \over \epsilon} \sinh x_0 \right) + 2\log g_b \; .
\ee 
This is smallest (and thus gives the correct result) for $\cosh t > e^{6 \log g_b \over c}  \sinh x_0$. 

%\subsubsection{Relation between the calculations}

In the next section, we will consider the direct CFT calculation of the entanglement entropies for the situations we have just described. As we explain in Section \ref{sec:bh-apps}, while the three calculations correspond to rather different physical scenarios, the underlying CFT calculation of the entanglement entropies is are directly related.

\section{BCFT calculation of entanglement entropies}

In this section, we move on to our central task: 
performing a direct CFT calculation of entanglement entropy for one or more intervals in the vacuum state of a BCFT on a half-line, for the thermofield double state of two BCFTs, or for the CFT state $|b ,  \tau_0 \rangle$ generated by a Euclidean BCFT path integral. We will argue that with certain assumptions, we can directly reproduce the holographic results described in the previous section.

\subsection{Entanglement entropy from correlation functions of twist operators}
\label{sec:ee-from-twists}

We begin by briefly recalling the CFT calculation of entanglement entropy (for more details, see
%a more detailed discussion, see 
\cite{Cardy:2007mb}). We consider a CFT or BCFT on a spatial geometry $M$ in some state $|\Psi \rangle$, defined by a Euclidean path integral on a geometry $H$ with boundary $M$. We would like to calculate the entanglement entropy $S_A = -\tr(\rho_A \log \rho_A)$ for a region $A \subset M$.

The entanglement entropy can be obtained from a limit of $n$-\emph{R\'{e}nyi entropies} $S^{(n)}_A$:
\begin{equation}
\label{eq:eedef}
S_A = \lim_{n\to 1}S^{(n)}_A\;, \quad
S^{(n)}_A := \frac{1}{1-n}\log \mbox{Tr}[\rho^n_A]\;.
\end{equation}
The matrix elements $\langle \phi^-_A| \rho_A|\phi^+_A\rangle$ are calculated from the path integral on a space $(\bar{H} H)_A$ formed from gluing two copies of $H$ along the complement of $A$ in $M$,\footnote{More precisely, the path integral corresponding to the second copy is the one associated with $\langle \Psi|$; any complex sources in the action should be conjugated.} where we set boundary conditions $\phi(x,\tau = \pm \epsilon) = \phi^\pm_A$ on either side of a cut $A$. The proper normalization is obtained by dividing by the same path integral without a cut along $A$.

\begin{figure}[t]
  \centering
  \includegraphics[scale=0.57]{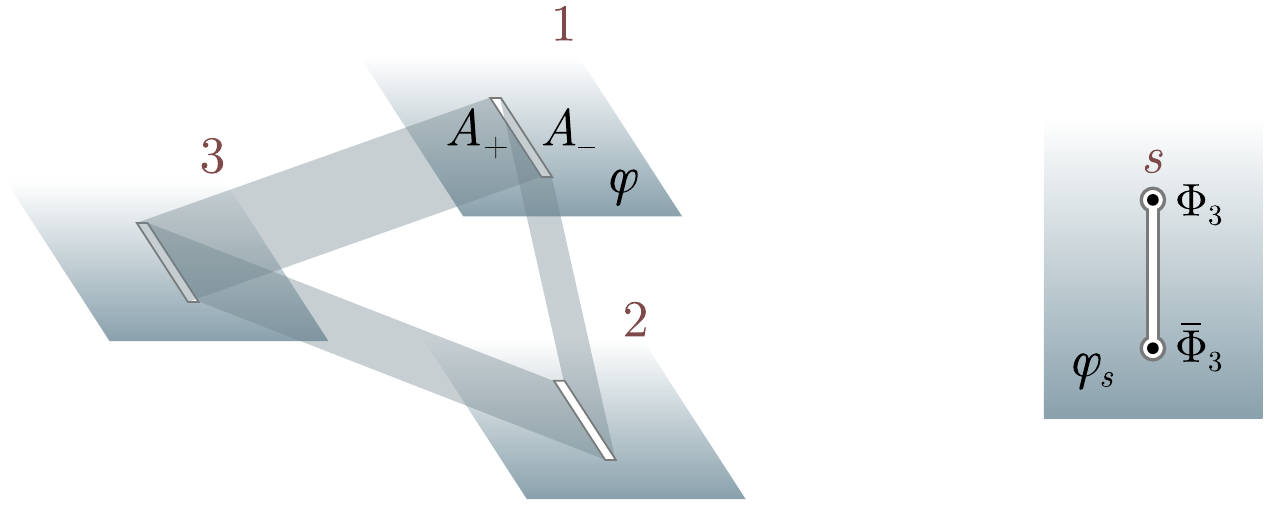}
  \caption{\emph{Left}. Three-replica geometry, $\mathcal{R}_3$, with a
    local field $\varphi$. \emph{Right}. Individual copies $s$, with
    boundary conditions for $\varphi_i$ implemented by twists $\Phi_3, \bar{\Phi}_3$.}
\label{fig:twist}
\end{figure}

The trace $\mbox{Tr}[\rho^n_A]$ is then obtained by the path integral on a \emph{replica geometry} $\mathcal{R}_n$  obtained by gluing $n$ copies of $(\bar{H} H)_A$ across the cut A, with the lower half of the cut on each copy glued to the upper half of the cut on the next copy, as shown in Figure \ref{fig:twist}. Including the proper normalization in the path integral expression for the density matrix gives
\begin{equation}
  \label{eq:42}
  \mbox{Tr} [\rho_A^n] = \frac{Z_n}{Z_1^n}\;,
\end{equation}
where $Z_n$ is the partition function for the CFT on $\mathcal{R}_n$.

The ratio $Z_n/Z_1^n$ can be expressed as a correlation function of \emph{twist operators} for a CFT/BCFT defined to be the product of $n$ copies of the original theory. A twist operator $\Phi_n(z)$ inserted at $z$ is defined via the path integral by inserting a branch cut ending at $z$, across which the fields in the $k$th copy of the (B)CFT are identified with fields in the $(k+1)$-st copy as we move clockwise around the branch point. Similarly, an anti-twist operator $\bar{\Phi}_n(z)$ inserts a branch cut ending at $z$ across which fields in the $k$th copy of the CFT/BCFT are identified with fields in the $(k-1)$-st copy as we move clockwise around the branch point.

\begin{figure}[t]
  \centering
  \includegraphics[scale=0.4]{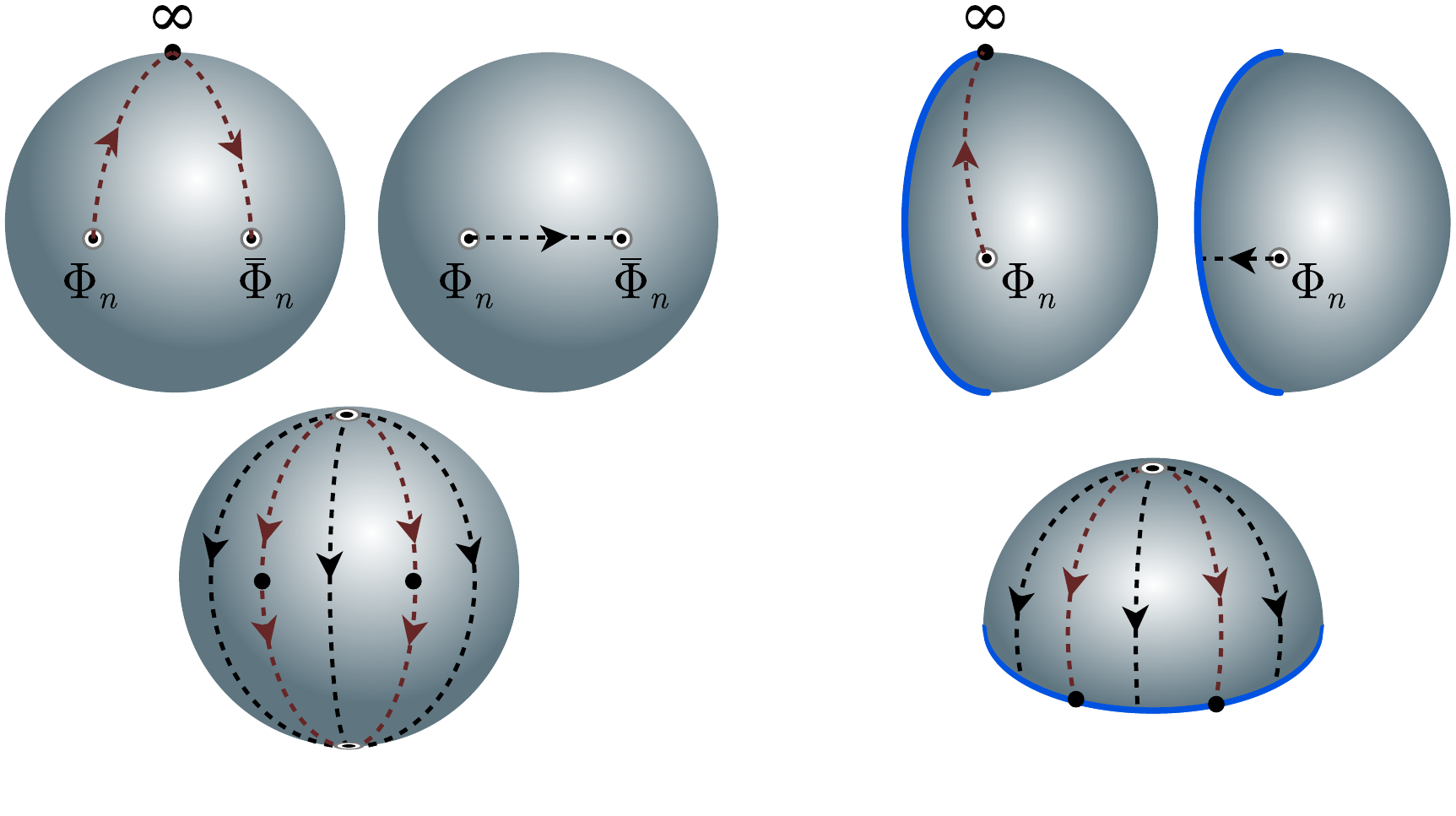}
  \caption{\emph{Left}. 
  Deforming the contour of the fundamental domain of $\mathcal{R}_n$ for a CFT.
  \emph{Right}.
  Performing the equivalent deformation on $\mathcal{R}_n$ for a BCFT.
  }
\label{fig:domain}
\end{figure}

In a CFT, every twist operator must come with an anti-twist operator, with the branch cut running between the two. For a BCFT, in contrast, we can have an \emph{unpaired} twist operator, with the branch cut running between the operator insertion and the boundary.
For both the CFT and BCFT, deforming the branch cut simply corresponds to changing the fundamental domain of the replica Riemann surface $\mathcal{R}_n$, as in Figure \ref{fig:domain}.
%This is related to the fact that, as we discuss below, the UV regularization of twists also inserts a boundary \cite{Cardy2016}.

\subsubsection*{Two-point function of the twist operators in a CFT}

%Then for example, 
The correlator $\langle \bar{\Phi}_n(x_1) \Phi_n(x_2) \rangle$ for the $n$-copy CFT on the real line with a branch cut running between $x_1$ and $x_2$ exactly computes the right-hand side of (\ref{eq:42}) for the case where $A$ is the interval $[x_1,x_2]$. The two point function takes a simple form, since as shown in \cite{Calabrese2009}, the twists fields $\Phi_n$, $\bar{\Phi}_n$ act like scalar primaries with scaling dimension
%with dimension
\begin{equation}
  \label{twist-scaling}
%  \langle \Phi_n(a,0)\bar{\Phi}(b,0)\rangle = |a-b|^{-2d_n}, \quad
d_n := \frac{c}{12}\left(n-\frac{1}{n}\right)
%h_n := \frac{c}{24}\left(n-\frac{1}{n}\right)
\end{equation}
and weights $h_n = \hat{h}_n = d_n/2$. Thus, we have
\be
\langle \bar{\Phi}_n(x_1) \Phi_n(x_2) \rangle \sim |x_1-x_2|^{-2 d_n} \; ,
\ee
as we will derive again below. 

To say more
about the coefficient, we need to define the twist operators more precisely by specifying the behavior of the CFT at the branch points. As a specific regularization, we can consider instead the $n$-copy theory defined on a space obtained by removing a disk of radius $\epsilon$ centered at each branch point and placing boundary condition labelled by $a_i$ at the $i$th resulting circular boundary \cite{Cardy2016}.\footnote{It will be convenient for our discussion below to allow different boundary conditions to regulate the different twist operators, but generally we can choose the same one for each.} The resulting path integral geometries for $Z_n$ (and $Z_1$) are then smooth. A conformal transformation
\be
z \mapsto  i \log\left({z - x_2 \over z -x_1}\right)
\ee
maps the original plane to a cylinder defined by the complex plane with identification $z \sim z + 2 \pi$, and the branch cut $[x_1,x_2]$ mapping to $\Re(z) = \pi$. 

For small $\epsilon$, the boundaries surrounding the branch points map to $z = \pm i \log[(x_2-x_1)/\epsilon]$ up to corrections of order $\epsilon$. Thus, the path integral geoemtry is a cylinder of length $\tau = 2 \log[(x_2-x_1)/\epsilon]$, with boundary condition $a_1,a_2$ at the two ends. The replica geometry is defined by gluing $n$ copies of this cylinder along the vertical branch cut, so corresponds to a cylinder with circumference $2 \pi n$. We can write the path integral on this space using boundary states as\footnote{We recall that the boundary state was defined using a circle of length 1. Scaling the cylinder to have this circumference, the length becomes $\tau /2 \pi n$.}
\be
\label{PF}
Z_n = \langle a_2| e^{- {\tau \over 2 \pi n} H} | a_1 \rangle
\ee
where $H$ is the Hamiltonian for the CFT on a circle of unit length. For large $\tau$, the operator inside approaches a projector to the vacuum state
\be
e^{- {\tau \over 2 \pi n} H} \to e^{- {\tau \over 2 \pi n} E_0} |0 \rangle \langle 0 | \; ,
\ee
where $E_0 = - \pi c / 6$ is the vacuum energy for a CFT on a circle of unit length. 

Thus, we get
\be
(Z_n)_{\epsilon \to 0} =  \left({|x_2 - x_1| \over \epsilon}\right)^{c \over 6n} \langle a_2|0 \rangle \langle 0| a_1 \rangle \; .
\label{eq:Z_n}
\ee
Finally,
\be
\label{CPhiPhi}
\langle \bar{\Phi}_n(x_1) \Phi_n(x_2) \rangle = {Z^{a,\epsilon}_n \over \left(Z^{a,\epsilon}_1\right)^n} = (\langle a_2|0 \rangle  \langle 0 | a_1 \rangle)^{(1-n)} \left({|x_1-x_2| \over \epsilon}\right)^{-2 d_n} \; .
\ee
Making use of this in (\ref{eq:eedef}) and (\ref{eq:42}) gives the standard result for the entanglement entropy of an interval. With our original definition of $\epsilon$, we have $c/3 \log(L/ \epsilon) +  \log g_{a_1} + \log g_{a_2}$, however, it will be convenient to take $a_1 = a_2 = a$ and absorb the last two terms here into the definition of $\epsilon$.

\subsubsection*{One-point function of the twist operator on a half space}

In a BCFT, the twist operators also have a non-vanishing one-point function, related to the R\'{e}nyi entropies for an interval $[0,x]$ in the vacuum state of the BCFT on a half-space $x \ge 0$.
%including the boundary for the vacuum state of a BCFT defined on a half-space $x \ge 0$. 
We can calculate this using the regularization defined above.

%It will be convenient to consider the 
We will consider a BCFT on the UHP with boundary condition $b$ at $\Im(z) =0$, with the twist operator at $z_1 = x_1 + i y_1$ regulated by boundary condition $a$. The conformal transformation
\be
z \mapsto  i \log\left({z - z_1^* \over z - z_1}\right)
\ee
maps the upper half-plane to a cylinder defined by the complex plane with identification $z \sim z + 2 \pi$, where the boundary along the real axis maps to the interval $[0, 2 \pi]$ on the real axis. The circle of radius $\epsilon$ regulating the twist operator maps (in the limit of small $\epsilon$) to a second end of the cylinder at $\Im(z) = \log(2 y_1 / \epsilon)$. 

Thus, the one-point function is $Z_n/(Z_1)^n$, where $Z_n$ is the partition function on a cylinder of circumference $2 \pi n$ and height $ \tau = \log (2 y_1 / \epsilon)$. Using the second equation in (\ref{CPhiPhi}), we have that
\be
\label{BPhi}
\langle \Phi_n(z_1, \bar{z}_n) \rangle = {Z_n \over (Z_1)^n} = (\langle a|0 \rangle  \langle 0 | b \rangle)^{(1-n)} \left|{2 y_1 \over \epsilon}\right|^{-d_n} \; ,
\ee
Interpreting the $\Re(z)$ direction as Euclidean time, this gives $\tr(\rho^n)$ for an interval $[0,y_1]$ in the vacuum state of a BCFT on a half-space. From (\ref{eq:eedef}), the entanglement entropy associated with this density matrix is
\be
S =  {c \over 6} \log \left({2 y_1 \over \epsilon}\right) + \log(g_a) + \log(g_b) \; .
\ee
The term $\log(g_a)$ in the regulator can be absorbed by a redefinition of $\epsilon$ to give the result (\ref{Sint}).\footnote{Note that this is the same redefinition as the previous subsection.} On the other hand, boundary entropy term $\log(g_b)$ is physical. It is equal to the difference between the BCFT entanglement entropy and the half the entanglement entropy in the parent CFT for an interval of length $2y_1$, with twist regularization fixed.

\subsection{Two-point function of twist operators on a half-space}

We are now ready for our main calculation. We consider the correlator on the UHP
of a twist operator at $z_1$ and an anti-twist operator at $z_2$.
%, required to compute the entanglement entropy for an interval in the vacuum state of a BCFT on a half-space.
As discussed in Section \ref{sec:bcft}, 
we can express the two-point function here either as a sum of bulk one-point functions (the {\it bulk channel}), or as a sum of boundary two-point functions (the {\it boundary channel}), using the bulk OPE or the BOE respectively. We now consider these expressions explicitly.

\subsubsection*{Boundary channel for the two-point function}

%As described in section 2, 
The boundary channel for the BCFT two-point function is obtained by first expanding each operator using the BOE, so that the bulk two-point function becomes a sum of boundary two-point functions. The contribution of two-point functions involving all the operators in a multiplet of the Virasoro symmetry sums to a conformal block. Using the general result (\ref{BCFT2pt}) with (\ref{2ptBoundary}), we find % have that
\be
\label{boundary2pt}
\uavg{ \Phi_n(z_1, \bar{z}_1) \bar{\Phi}_n (z_2, \bar{z}_2)} = \left[ {\eta \over 4 y_1 y_2} \right]^{d_n} \sum_I \mathcal{B}^{b}_{\Phi I} \mathcal{B}^{b}_{\bar{\Phi} I} \mathcal{F}(c,\Delta_I ,{d_n \over 2}| \eta)
\ee
where 
\begin{equation}
\eta = \frac{(z_1 - \bar{z}_1)(z_2 - \bar{z}_2)}{(z_1 - \bar{z}_2)(z_2 - \bar{z}_1)}\;.
\end{equation}
and where $I$ indexes untwisted boundary operators in the $n$-fold product theory. As we review in Appendix B, the BOE coefficients here can be expressed in terms of correlators of boundary operators in the original BCFT.\footnote{To avoid cluttering our notation further, we will generally leave $n$ implicit in our BOE coefficients $\mathcal{B}$.}
Writing the cross ratio explicitly in terms of real coordinates, we have
\begin{align}
  \label{uhp-eta-1}
  \eta & %= \frac{z_{1\bar{1}}z_{2\bar{2}}}{z_{1\bar{2}}z_{2\bar{1}}}
   =
  \frac{4y_1y_2}{(x_1-x_2)^2 + (y_1+y_2)^2} = 1 -
  \frac{(x_1-x_2)^2+(y_1-y_2)^2}{(x_1-x_2)^2+(y_1+y_2)^2}\;,
\end{align}
so we see that $\eta$ is a real number in $[0,1]$, with $\eta \to 1$ in the limit where $z_1$ and $z_2$ are much closer to each other than the boundary  and $\eta \to 0$ in the limit where $z_1$ and $z_2$ are much closer to the boundary than to each other.

Consider
the contribution from the term where only the boundary identity operator is kept in each BOE (\ref{boe-2}).  
This is equal to the disconnected term in the two-point function that factorizes into the product of one-point functions, and hence
\bea
\label{boundarychannelID}
\langle \Phi_n(z_1, \bar{z}_1) \bar{\Phi}_n (z_2, \bar{z}_2)\rangle_{\text{UHP},\bf 1}^b &=& {\mathcal{B}^{b}_{\Phi \bf 1} \mathcal{B}^{b}_{\bar{\Phi} \bf 1}  \over [4 y_1 y_2]^{d_n}} \cr
&=& {g_b^{2(1-n)} \epsilon^{2 d_n} \over [4 y_1 y_2]^{d_n}} \; ,
\eea
%where we have kept only the identity term in the BOEs (\ref{boe-2}) for each operator and 
where we have read off $\mathcal{B}^{b}_{\Phi \bf 1}$ from (\ref{BPhi}). In general, this contribution should dominate the correlator in the limit $\eta \to 0$, where the two operators approach the boundary. 

\subsubsection*{Bulk channel for the two-point function}

We can obtain an alternative expression for the two-point function using the bulk OPE to express the product $\Phi_n (z_1,\bar{z}_1)\bar{\Phi}_n(z_2,\bar{z}_2)$ as a sum of bulk operators. This reduces the two-point function to a sum of one-point functions.

Using the general result (\ref{BCFT2pt}) with (\ref{2ptBulk}) for this  bulk-channel expression for the two-point function, we obtain
\bea
\label{bulk2pt}
\uavg{\Phi_n(z_1, \bar{z}_1) \bar{\Phi}_n (z_2, \bar{z}_2)} &=& \left[ {\eta \over 4 y_1 y_2} \right]^{d_n} \sum_i C^i_{\Phi_n \bar{\Phi}_n} \mathcal{A}^b_i \mathcal{F}(c,h_i , {d_n \over 2} | 1 - \eta) \cr\; 
&=& \left[ {1-\eta \over |z_1 - z_2|^2} \right]^{d_n} \sum_i C^i_{\Phi_n \bar{\Phi}_n} \mathcal{A}^b_i \mathcal{F}(c,h_i , {d_n \over 2} | 1 - \eta) \; ,
\eea
where $i$ indexes untwisted operators in the $n$-fold product CFT. 
Again, it will be useful below to note the contribution where we keep only the bulk identity operator term in the OPE (\ref{ope}): %, which gives
\bea
\label{bulkchannelID}
\langle \Phi_n(z_1, \bar{z}_1) \bar{\Phi}_n (z_2, \bar{z}_2)\rangle_{\text{UHP},\bf 1}^b &=& {C^{\bf 1}_{\Phi_n \bar{\Phi}_n} \mathcal{A}^b_{\bf 1} \over |z_1 - z_2|^{2 d_n}} \cr
&=& {\epsilon^{2 d_n} \over |z_1 - z_2|^{2 d_n}} \; ,
\eea
where we have used $\mathcal{A}^b_{\bf 1} = 1$ and $C^{\bf 1}_{\Phi_n \bar{\Phi}_n} = \epsilon^{2 d_n}$ from (\ref{CPhiPhi}). This contribution should dominate the correlator in the limit $\eta \to 1$, where the two operators approach each other away from the boundary. %are much closer to each other than to the boundary.

\subsection{R\'{e}nyi entropy}
\label{sec:renyi}

We now use our results to calculate the Renyi entropy for an interval $A = [y_1,y_2]$ for the vacuum state of a BCFT on a half space $y > 0$. This is related to the two-point function of twist operators on the upper half-plane as 
\be
e^{(1-n)S_A^{(n)}} = \uavg{ \Phi_n(z_1,\bar{z}_1)\bar{\Phi}_n(z_2,\bar{z}_2)}\;,
\ee
where we take $z_1 = (0,y_1)$ and $z_2 = (0,y_2)$. 

\subsubsection*{Bulk and boundary limits}

First, consider the R\'{e}nyi entropy in the limits $\eta \to 0$ and $\eta \to 1$, where the twist operator two-point function is given by (\ref{boundarychannelID}) and (\ref{bulkchannelID}) respectively. In this case, we find that
\be
S_A^{(n)} = \left\{ \ba{ll}
\displaystyle{c \over 12} {n+1 \over n}\log \left( {2 y_1 \over \epsilon} \right) + {c \over 12} {n+1 \over n}\log \left( {2 y_2 \over \epsilon} \right)  + 2 \log g_b & \qquad \eta \to 0 \cr \displaystyle
 {c \over 6} {n+1 \over n} \log \left( {|y_2 - y_1| \over \epsilon} \right) & \qquad \eta \to 1 \; .\ea \right.
\ee
Taking the $n \to 1$ limit, these give entanglement entropies
\be
\label{vacresults}
S_A = \left\{ \ba{ll} 
\displaystyle{c \over 6} \log \left( {2 y_1 \over \epsilon} \right) + {c \over 6} \log \left( {2 y_2 \over \epsilon} \right)  + 2 \log g_b & \qquad \eta \to 0 \cr\displaystyle{c \over 3} \log \left( {|y_2 - y_1| \over \epsilon} \right) & \qquad \eta \to 1\;. \ea \right.
\ee
We see that these precisely match the holographic results (\ref{disc}) and (\ref{conn}). 

The result (\ref{disc}) is expected to be valid for any holographic CFT in some finite interval around $\eta = 0$ where the RT surface is disconnected, while the result (\ref{conn}) is expected to be valid in a finite interval around $\eta = 1$ in the case where the holographic theory can be modelled by a purely gravitational ETW brane. Thus, the results (\ref{vacresults}) have a much larger range of validity than we would naively expect from the CFT point of view. We would now like to understand from the CFT perspective how this larger range of validity for the vacuum results can arise.

\subsubsection*{Entropies at large $c$}
\label{sec:semiclassical-blocks}

We begin with the general
expressions (\ref{boundary2pt}) and (\ref{bulk2pt}) for the twist operator two-point function. General closed-form expressions for the conformal blocks are not known, but in the \emph{semiclassical limit} $c
\to \infty$, the chiral conformal blocks %in (\ref{boundary2pt}) and (\ref{bulk2pt}) 
exponentiate \cite{Belavin1984}:\footnote{There is a beautiful but  non-rigorous argument for exponentiation from Liouville theory,  using the explicit structure constants \cite{Dorn1994,
    Zamolodchikov1995} and the path integral. We refer the interested reader to
the clear account in \cite{Harlow2011a}.}
\begin{align}
  \label{semi-exp}
  \mathcal{F}(c, h_{\text{int}} , h | \eta) & \overset{c\to\infty}{=}
  \exp\left[-\frac{c}{6}f\left(
                                                      \frac{h_{\text{int}}}{c},\frac{h}{c}, \eta\right)\right] \;.
\end{align}
The exponent $f$ is called the \emph{semiclassical block}.\footnote{In general, this depends on the set of external weights, but our notation takes into account that all of the external weights are identical.} In our case of identical external weights, recursion
relations for the block allow one to commute the limits $c \to \infty$
and $h_{\text{int}}/c,h/c \to 0$ \cite{Zamolodchikov1987, Hartman2013}.
Hence, the semiclassical blocks associated to \emph{light} internal operators $h_{\text{int}} = O(c^0)$ are just the vacuum
(semiclassical) block:
\be
f_0\left(\frac{h}{c}, \eta\right) \equiv f\left(0,\frac{h}{c}, \eta\right) \; .
\ee
We can apply these results to our two-point function of twist operators, for which all of the external dimensions are $d_n/2$, and the central charge of the replicated CFT is $nc$. 

We find that the $c \to \infty$ limit of the expressions (\ref{boundary2pt}) and (\ref{bulk2pt}) for the twist operator two-point function in the boundary and bulk channels become
\begin{align}
\uavg{
  \Phi_n(z_1,&\bar{z}_1)\bar{\Phi}_n(z_2,\bar{z}_2)}
   \notag \\ & = \left( {\eta \over 4 y_1 y_2} \right)^{d_n} \left[\hat{\mathcal{D}}_{\text{L}} e^{-\frac{nc}{6}f_{\hat{0}}\left(\frac{d_n}{2nc},\eta\right)}  +
\sum_{J_\text{H}} \Bcal^b_{\Phi J}\Bcal^b_{\bar{\Phi}
    J}e^{-\frac{nc}{6}f\left(\frac{\hat{\Delta}_{J}}{nc},\frac{d_n}{2nc},  \eta\right)} \right] \label{bound-semiclassical}\\
  & = \left( {\eta \over 4 y_1 y_2} \right)^{d_n} \left[\mathcal{D}_{\text{L}} e^{-\frac{nc}{6}f_{0}\left(\frac{d_n}{2nc},1-\eta\right)} +
    \sum_{j_\text{H}}  C^j_{\Phi_n \bar{\Phi}_n} \mathcal{A}^b_j
    e^{-\frac{nc}{6}f\left(\frac{h_j}{nc} ,\frac{d_n}{2nc}, 1-\eta\right)} \right]\label{bulk-semiclassical},
\end{align}
where $J_\text{H}, j_\text{H}$ range over heavy internal operators,
and $\hat{\mathcal{D}}_{\text{L}}$ and $\mathcal{D}_{\text{L}}$ are degeneracy factors multiplying the vacuum channel:
\begin{equation}
  \label{eq:30}
  \hat{\mathcal{D}}_\text{L} = \sum_{J_\text{L}} \Bcal^b_{\Phi J}\Bcal^b_{\bar{\Phi} J},
                               \quad  \mathcal{D}_\text{L} = \sum_{j_\text{L}}  C^j_{\Phi_n \bar{\Phi}_n} \mathcal{A}^b_j\;.
\end{equation}

As $c\to \infty$, the sums (\ref{bound-semiclassical}) and
(\ref{bulk-semiclassical}) should be dominated by the exponential with smallest exponent, if the coefficients of the exponential in the sum are not too large. 
More precisely, let us now make two assumptions:
\begin{enumerate}
    \item The contribution of all heavy internal operators, in a neighbourhood around $\eta=0$ or $\eta=1$ in the respective  channel, is exponentially suppressed in $c$. We will take heavy to mean any operators whose dimension scales as $O(c)$ or greater. 
    \item The degeneracy factors $\hat{\mathcal{D}}_\text{L},\mathcal{D}_\text{L}$ are given by the vacuum contribution times some multiplicative correction that does not change the leading exponential in $c$ behaviour.
\end{enumerate}

If the neighborhoods described in the first assumption meet at some point $\eta_*^n$, so that they cover the entire interval $\eta \in [0,1]$, we can conclude that large-$c$ behaviour of the correlator is given by the larger of the vacuum block contribution in the boundary channel or the vacuum block contribution in the bulk channel for the entire interval $\eta \in [0,1]$. 
This behaviour is  commonly known as {\it vacuum block dominance}. 

Under our first assumption of vacuum block dominance, %we have that the 
the R\'{e}nyi entropy for an interval $[y_1,y_2]$ is given by
\begin{small}
\be
S_A^{(n)} = \left\{ \ba{ll} \displaystyle{c \over 6} {n+1 \over n} \log (y_1 + y_2)  + {c \over 6} {n \over n-1} f_0 \left({1 \over 24}\left(1 - {1 \over n^2}\right), {4 y_1 y_2 \over (y_1 + y_2)^2 } \right) + {1 \over 1-n} \log \hat{\mathcal{D}}_\text{L}& \quad \eta < \eta_*^n \cr \displaystyle
{c \over 6} {n+1 \over n} \log (y_1 + y_2)  + {c \over 6} {n \over n-1} f_0 \left({1 \over 24}\left(1 - {1 \over n^2}\right), {(y_1 - y_2)^2 \over (y_1 + y_2)^2 } \right) + {1 \over 1-n} \log \mathcal{D}_\text{L}\;,& \quad \eta > \eta_*^n \ea \right.
% S_A^{(n)} = \left\{ \ba{ll} \displaystyle{c \over 6} {n+1 \over n} \log (y_1 + y_2)  + {c \over 6} {n \over n-1} f_0 \left({1 \over 24}\left(1 - {1 \over n^2}\right), {4 y_1 y_2 \over (y_1 + y_2)^2 } \right) + {1 \over 1-n} \log \hat{\mathcal{D}}_\text{L}& \quad \eta < \eta_*^n \cr \displaystyle
% {c \over 6} {n+1 \over n} \log (y_1 + y_2)  + {c \over 6} {n \over n-1} f_0 \left({1 \over 24}\left(1 - {1 \over n^2}\right), {(y_1 - y_2)^2 \over (y_1 + y_2)^2 } \right) + {1 \over 1-n} \log \mathcal{D}_\text{L}\;,& \quad \eta > \eta_*^n \ea \right.
\ee
\end{small}
$\!\!\!$where $\eta^n_*$ is the value of $\eta$ at which the lower expression becomes larger than the upper one. In the limit $n \to 1$, %that gives the entanglement entropy, 
the behavior of the semiclassical vacuum block follows from the result that for small $\alpha = (n-1)/12$ \cite{Hartman2013},
\be
f_0(\alpha, \eta) = 12 \alpha \log \eta + {\cal O}(\alpha^2) \; ,
\ee
as we derive in Section \ref{sec:bcft-mult}.

Under our second assumption of vacuum block dominance, we have that (at order $c$)
\beas
\lim_{n \to 1} {1 \over 1-n} \log \hat{\mathcal{D}}_\text{L} &=& \lim_{n \to 1} {1 \over 1-n} \log (\mathcal{B}^b_{\Phi \bf 1} \mathcal{B}^b_{\bar{\Phi} \bf 1} ) \cr
&=& -{c \over 3} \log  \epsilon + 2 \log g_b \cr
\lim_{n \to 1} {1 \over 1-n} \log \mathcal{D}_\text{L} &=& \lim_{n \to 1} {1 \over 1-n} \log (C^{\bf 1}_{\Phi_n \bar{\Phi}_n} \mathcal{A}^b_{\bf 1} ) \cr
&=& -{c \over 3} \log  \epsilon 
\eeas
up to contributions $O(c^0)$. 
Note that, by keeping the boundary entropy term, we are assuming that it, too, is $O(c)$. 

Using these results and the results for the semiclassical blocks gives 
\be
\label{Sresults}
S_A = \lim_{n \to 0} S_A^{(n)} = \left\{ \ba{ll} \displaystyle{c \over 6} \log \left({2 y_1 \over \epsilon}\right) + {c \over 6} \log\left({2 y_2 \over \epsilon}\right)  + 2 \log g_b& \qquad \eta < \eta_* \cr \displaystyle
{c \over 3}\log\left({|y_2 - y_1| \over \epsilon}\right) & \qquad \eta > \eta_* \ea \right.
\ee
where $\eta_*$ is the value of $\eta$ where the two expressions coincide. These are exactly the results (\ref{vacresults}) we obtained keeping only the contributions from boundary and bulk identity operators.
Thus, we see that the assumption of vacuum block dominance provides the extended range of validity for the formulas in (\ref{vacresults}), so that the results match our gravitational calculation with a purely gravitational ETW brane.

\subsection{BCFT requirements for vacuum block dominance}

Our expression in (\ref{Sresults}) now matches precisely the gravitational calculation, \eqref{disc} and \eqref{conn} for all $\eta$, at leading order in $c$. 
Following the previous work for bulk CFTs \cite{Hartman2013,Perlmutter2014}, let us now explore what constraints our vacuum block dominance assumptions place on the spectrum and OPE data of the BCFT.  

\subsubsection{Boundary channel}

We begin with the disconnected phase in the boundary channel that dominates in a neighbourhood of $\eta =0$. 
Our first assumption held that the contribution of heavy boundary operators was exponentially suppressed in $c$ and does not contribute at leading order. 
We will examine this claim in a cascading series of steps, from heaviest to lightest operators.

First, looking at operators of dimension $O(c^\alpha)$ for $\alpha> 1$, we find that agreement with the gravity calculation seems to place rather weak constraints on the BCFT. In particular, the convergence of the boundary OPE can be used in an exactly analogous manner to the convergence of the bulk OPE \cite{Pappadopulo:2012jk} to show that the contribution of all operators of dimension $\hat \Delta > O(c)$ is exponentially suppressed in the central charge. 

We then need only worry about operators up to dimension $O(c)$.  Define $\rho_{b,n}(\delta)\, \D \delta$ to be the number of untwisted $n$-fold product boundary operators with dimensions $\hat \Delta \in c[\delta, \delta + \D \delta]$, and define a measure of the average twist-operator BOE coefficients by
\begin{equation}
\label{Bdef}
  |B_n(\delta)|^2 = \frac{ \sum_{\hat\Delta_I \in c[\delta, \delta + \D \delta]} \bar\Bcal^b_{\Phi I} \bar\Bcal^b_{\bar{\Phi} I}}{\sum_{\hat\Delta_I \in c[\delta, \delta + \D \delta]} 1 }\;.  
\end{equation}
where we have introduced $\bar\Bcal^b_{\bar{\Phi} I} = \epsilon^{-d_n} g_b^{n-1} \Bcal^b_{\bar{\Phi} I} $ to remove a universal prefactor that appears in all the BOE coefficients (see Appendix B). 
We can use the known small $\eta$ expansion of the semiclassical block \cite{Hartman2013},
\be
f(h_{\text{int}},h_{\text{ext}},\eta) = 6 (2h_{\text{ext}} - h_{\text{int}}) \log \eta - {h_{\text{int}} \over 2} \eta + {\cal O}(\eta^2)\;,
\ee
to write the bracketed expression in (\ref{bound-semiclassical}) as
\be
e^{-{nc \over 12}\left(1 - {1 \over n^2}\right) \log \eta/\epsilon +2(1-n)\log g_b} \int^{O(1)}_0 \D \delta \,\rho_{b,n}(\delta) |B_n(\delta)|^2  e^{c \delta \log \eta + c \delta \eta + {\cal O}(\eta^2)}\;.
\ee
In this expression, the heavy operators will not contribute to the order $c$ entanglement entropy if the integral over of any region bounded away from zero is exponentially suppressed in $c$ as compared to the integral near zero.  
This constrains the product of the density of operators appearing in the twist OPE and their OPE coefficients so as not to grow so quickly as to overcome the suppression from the block. 
For $\eta \ll 1$, this requires 
\begin{equation}
    \log \left( \rho_{b,n}(\delta) |B_n(\delta)|^2 \right)  < c \delta \log( \eta^{-1}) \quad \mathrm{for} \quad \delta \gtrsim 0 \; .
\end{equation}
In particular, requiring the CFT calculation to agree with the gravity result in an interval $0 < \eta < \hat{\eta} \ll 1$ implies that $\rho_{b,n}(\delta) |B_n(\delta)|^2$ grows more slowly than $\exp(c \delta \log (1/\hat{\eta})$. Extending to a larger range with $\hat{\eta}$ not necessarily much less than 1 gives a stronger constraint, but the exact form requires more detailed knowledge of the semiclassical block.
% \footnote{Note that  we can let the measure grow like
% \begin{equation}
%      \rho_{b,n}(\delta) |B_n(\delta)|^2  = \exp \left( c \beta(\delta) \right)
% \end{equation}
% for any bounded function $\beta(\delta)$ that is sub-linear in $c$, and our conclusion will hold when $\eta$ is small enough (as was noted in \cite{Perlmutter2014}). 
% This is not sufficient for agreement with the gravitational picture for all $\eta < \eta_*$.}

Let us then focus on the lower limit of this integral and consider only operators of dimension less than $O(c^\alpha)$ for $\alpha < 1$, where we can approximate the semiclassical block by the vacuum block for all operators, up to $O(c^{\alpha-1})$ corrections. The gravity calculation predicts that the leading exponential in $c$ behavior of the result matches the vacuum channel contribution, so we require that
\be
\label{DFcon}
\sum_{I_L} \bar\Bcal^b_{\Phi_n I} \bar\Bcal^b_{\bar{\Phi_n} I} 
\ee
is subexponential in $c$.\footnote{It is also interesting to consider the constraints on the CFT assuming that we have a conventional gravitational theory with a usual semiclassical expansion. In this case, the corrections to the entropy are expected to be of order $c^0$ (as opposed to some larger power of $c$ or $\log c$). In this case, we would obtain stronger constraints on the BCFT. However, for this paper, we focus on the constraints arising from demanding that the order $c$ terms in the entropies match with the classical gravity calculation.} In Appendix B, we recall that the coefficients $\bar\Bcal^b_{\Phi_n I}$ can be expressed in terms of $n$-point correlations functions of light boundary operators in the original BCFT, so this constraint can be translated into a constraint on the spectrum and $n$-point functions of the original BCFT. We consider the case $n=2$ in more detail below. 

%While this `light' region of the spectrum is best written as a sum, as we did in %\eqref{bound-semiclassical}, we will approximate it as the integral 
%\begin{equation}
%    e^{\frac{-nc}{6}f_0(\eta) + d_n \log \epsilon}\int^{c^{\alpha-1}}_\epsilon \D \delta %\,\rho_b(\delta) |B(\delta)|^2 \, .
%\end{equation}
%For our second numbered assumption in the previous subsection to be valid, we require that the %integral gives corrections to the coefficient of the semiclassical block that do not change %the leading $O(e^c)$ contribution. 
%his will hold provided
%\begin{equation}
%  \beta_n(c,\delta)  \lesssim 2(1-n) \log g_b \quad \mathrm{for} \quad \delta \lesssim c^{\alpha-1}\, ,
%\end{equation}
%that is, the spectrum and BOE coefficients are less than an expression exponential in $c$.

%For example, if we take $\beta_n(c,\delta)$ to have the form
%\begin{equation}
%    \beta_n(c,\delta) = b_{n,0} + O(\delta,c^{-1})
%\end{equation}
%then this bounds 
%\begin{equation}
%    b_{n,0} \lesssim 2(1-n) \log g_b \, .
%\end{equation}
%Or, if we let
%\begin{equation}
%    \beta_n(c,\delta) = \tilde b_{n,1} \Delta 
%\end{equation}
%and we push this constraint all the way to a cutoff  $\Delta = \Delta_0 = O(c)$, we find 
%\begin{equation}
%    \tilde b_{n,1} \lesssim \frac{ 2(1-n) \log g_b }{\Delta_0} \, ,
%\end{equation}
%where the exact bound is sensitive to where we cut the integral off (as one might expect having ignored the suppression that comes from the block itself for finite $\delta$).

\subsubsection{Bulk channel}

We can largely repeat the above analysis in the bulk channel. Again, we find only weak constraints on operators of dimension $O(c^\alpha)$ for $\alpha> 1$. 
The convergence of the bulk OPE can be used now precisely as in \cite{Pappadopulo:2012jk} to show that the contribution of all operators of dimension $\Delta > O(c)$ is exponentially suppressed in the central charge. 
We then need only worry about operators up to dimension $O(c)$.  

Define  $\rho_n(\delta) \,\D \delta$ to be the number of bulk untwisted $n$-fold product operators with dimensions $\Delta \in c[\delta, \delta + \D \delta]$ and 
\be
AC_{b,n}(\delta) = \frac{ \sum_{\Delta_i \in c [\delta, \delta + \D \delta]} \bar C^i_{\Phi \bar{\Phi}} \mathcal{A}^b_i}{\sum_{\Delta_i \in c[\delta, \delta + \D \delta]} 1 }\;,
\ee
where $\bar C^i_{\Phi \bar{\Phi}} = \epsilon^{-2 d_n} C^i_{\Phi \bar{\Phi}}$. 
Then in this channel, we have 
\be
e^{-{nc \over 12}\left(1 - {1 \over n^2}\right) \log (1-\eta)/\epsilon} \int^{O(1)}_0 \D \delta \,\rho_n(\delta) AC_{b,n}(\delta)  e^{c \delta \log (1-\eta) + c \delta (1-\eta) + {\cal O}((1-\eta)^2)}\;.
\ee
For heavy operators to not contribute to the order $c$ entanglement entropy when $1-\eta \ll 1$, we require
\begin{equation}
  \log \left( \rho_n(\delta) AC_{b,n}(\delta) \right)  <  c \delta \log(1-\eta)^{-1} \quad \mathrm{for} \quad \delta \gtrsim 0 \;.
\end{equation}
This is analogous to the boundary channel condition, but with $\eta \to 1 - \eta$.

For operators with dimension $O(c^\alpha)$ for $\alpha < 1$, assumption 2 must hold in order to match with from gravity with a purely gravitational ETW brane. This requires that for the light operators, the sum 
\begin{equation}
    \sum_{i_L} \bar C^i_{\Phi \bar{\Phi}} \mathcal{A}^b_i
\end{equation}
should be sub-exponential in $c$. 

\subsection{Constraints on holographic BCFTs}

We have now spelled out explicitly a set of conditions on a BCFT that will ensure that the direct BCFT calculation of entanglement entropy matches with the gravity results in the holographic model with a purely gravitational ETW brane. However, we recall that the disconnected phase result (\ref{disc}) is universally valid for any holographic BCFT. Assuming that entanglement entropy has such a disconnected phase for some interval $\eta \in [0, \eta_*]$, as it does for the simple model, suggests that vacuum block dominance should hold for any holographic BCFT in an interval $\eta \in [0, \eta_n]$, where the upper end of the interval may depend on the Renyi index $n$.

From the results in the previous subsection, this implies a constraint
\begin{equation}
\label{Bbound}
    \log \left( \rho_{b,n}(\delta) |B_n(\delta)|^2 \right)  < c \beta^* \delta + O(c^a) \; ,  \quad a < 1\;, 
\end{equation}
where the quantities in the left side were defined in (\ref{Bdef}) and the preceeding paragraph. Here our knowledge of the semiclassical block was not sufficient to fix the $O(1)$ coefficient $\beta^*$ in this bound.  In addition, we have a constraint (\ref{DFcon}) on the light operators. We take these bounds to be novel constraints on which BCFTs can possibly have a gravitational dual.

Although we found an analogous bound
\begin{equation}
       \log \left( \rho_n(\delta) AC_{b,n}(\delta) \right)  \lesssim  c \gamma^* \delta +  O(c^a) \; ,  \quad a < 1 
\end{equation}
in the bulk channel, this should not be viewed as a constraint on the boundary expectation values $\mathcal{A}^b_i$. 
%This is because, 
While the disconnected phase is universal and depends only on the boundary entropy, the connected phase depends on the gravitational background (e.g. whether we have backreacting scalars in the solution dual to the BCFT vacuum). 
The vacuum solution for the bulk CFT is unique, but in contrast, there is no unique gravitational solution consistent with the symmetries of the BCFT. 

A useful diagnostic for the non-universal behaviour of entropy and the bulk background is when light operators have large, $O(c)$, expectation values that backreact on the gravitational solution:
\begin{equation}
    \uavg{\mathcal{O}_i(x,y)} = \frac{\mathcal{A}^b_i}{(2y)^{\Delta}} \; , \quad \mathcal{A}^b_i \sim c\;.
\end{equation}
Consistency with the large-$c$ factorization in the bulk then implies there is a large family of ``multi-trace" operators of the schematic form $\mathcal{O}^m$ with expectation values $\uavg{\mathcal{O}^m} \sim c^m$. 
When calculating the twist correlation function, this tower of operators must be  resummed into a new semiclassical block, just as with the gravitational Virasoro descendants.
For a BCFT, the form of the semiclassical block is theory-dependent and hence non-universal. 

Thus, in the bulk channel vacuum-block dominance is not required by the theory. 
We must choose to restrict to those boundary states without semiclassical expectation values where non-universal contributions can be ignored.\footnote{The same limitation holds for previous bulk CFT calculations. When light bulk operators have large expectation values that backreact on the geometry, the entanglement entropy of a region is no longer universal and is not determined by vacuum block dominance.} 

\subsubsection{Constraints on the BCFT base theory}

The constraint (\ref{Bbound}) involves both the spectrum of boundary operators in the $n$-copy theory and the BOE coefficients for twist operators in this theory. As we review in appendix B, both of these can be related to the spectrum and OPE data for boundary operators in the single-copy BCFT; we can make use of these relations to convert the constraint (\ref{Bbound}) to a direct statement about the single-copy BCFT. 

In particular, consider the case of $n=2$, where the branched geometry (including a regulator boundary for the twist operator as above) is conformal to the annulus. 
The Virasoro primaries appearing in the the $n=2$ twist BOE, analogously to the bulk CFT case in \cite{Perlmutter2014,Chen:2013kpa,Calabrese:2010he}, contain products of base primaries of the form
\begin{equation}
\label{simple2}
     \mathcal{O}_I = \mathcal{O}_i \otimes \mathcal{O}_i \,,
\end{equation}
For these operators, as we show in appendix \ref{sec:app-twist-boe-coeff}, the BOE coefficients are 
\begin{equation}
    \bar\Bcal^b_{\Phi_2 I} \bar\Bcal^b_{\bar{\Phi_2} I} = 16^{-2\hat \Delta_i} \, ,
\end{equation}
identical to the bulk case in  \cite{Perlmutter2014,Chen:2013kpa,Calabrese:2010he} up to the non-standard normalization of the twist operators induced by the boundary. Taking into account only these primaries, we have a constraint from (\ref{DFcon}) that
\be
\label{constrL}
\sum_{i_L} 16^{-2\hat \Delta_i} \; ,
\ee
is sub-exponential in $c$, where the sum is over light boundary primary operators in the original BCFT. This will be true if the number of light boundary primaries in the base theory is also sub-exponential in $c$.

Note that the BOE also contains primaries composed of products of descendants in the base theory, such as 
\begin{equation}\label{eq:funny-descendant}
   \mathcal{O}_i \otimes L_{-1}^2 \mathcal{O}_i -2 \tfrac{h_i +1/2}{h_i} L_{-1} \mathcal{O}_i \otimes L_{-1}\mathcal{O}_i+ L_{-1}^2 \mathcal{O}_i \otimes \mathcal{O}_i  \,.
\end{equation}
These are primaries with respect to the orbifold Virasoro generators:
\begin{equation}
    L_{m} \otimes 1 + 1 \otimes L_{m} \, ,
\end{equation}
but are generated by even powers of the antisymmetric linear combinations
\begin{equation}
    L_{m} \otimes 1 - 1 \otimes L_{m} \, .
\end{equation}
To estimate the number of such primaries, we use the Hardy-Ramanujan Formula \cite{doi:10.1112/plms/s2-17.1.75}, which gives an asymptotic estimate for the number of descendants (partitions $p(k)$) at a given level $k$:
\begin{equation}
    p(k) \sim e^{2 \pi \sqrt{k/6}}\, .
\end{equation}
If the density of light primaries for a single BCFT is  sub-exponential, as above, including the contribution of the extra primaries in the 2-copy BCFT not of the form (\ref{simple2}) generates no new contributions exponential in $c$. Thus, we do not get a substantially stronger constraint from their inclusion.

\subsection{Black hole applications}
\label{sec:bh-apps}

We have seen that under the assumption of vacuum block dominance, the direct BCFT calculation of entanglement entropy for an interval in the vacuum state of the theory on a half-line matches exactly with the gravity calculation using an effective bulk theory with a purely gravitational ETW brane.

In this section, we show that essentially the same BCFT calculation allows us to reproduce the gravitational results (\ref{BSconn}) and (\ref{BSdisc}) for the entanglement entropy of an interval in a Euclidean time-evolved boundary state $|b, \tau_0 \rangle$, and the results (\ref{BRconn}) and (\ref{BRdisc}) for the entanglement entropy of the auxiliary radiation system in BCFT model of a two-sided black hole coupled to a radiation bath.

\begin{figure}[t]
  \centering
  \includegraphics[scale=0.25]{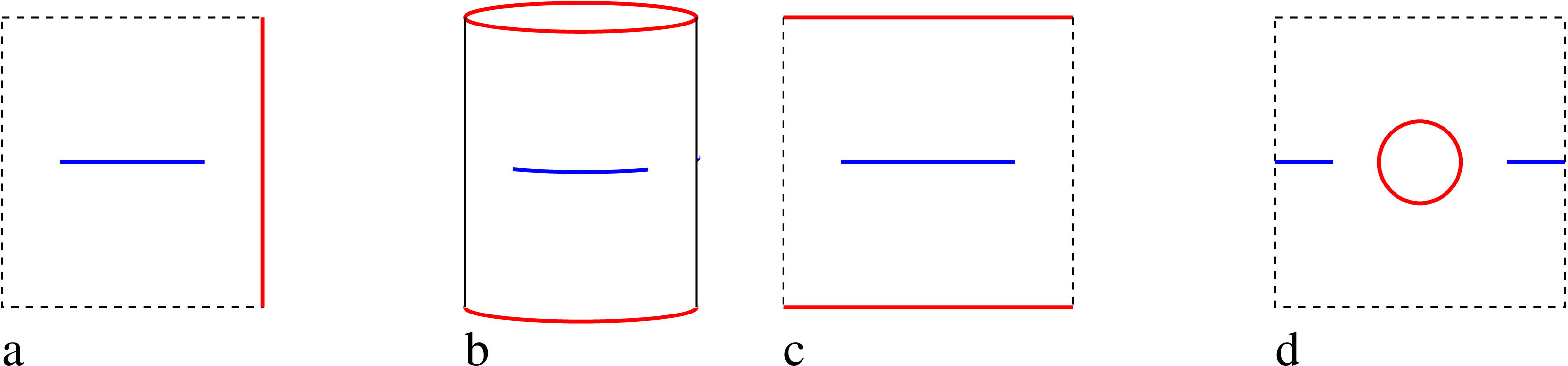}
  \caption{Relation between Euclidean path integrals for the different BCFT setups in which we are calculating entanglement entropy. In (a), we have the half-space $x>0$. In (b), we have the path integral preparing the boundary state $|b,\tau_0\rangle$. We decompactify from a circle to a line to obtain the global quench geometry (c), which is equivalent (under a global conformal transformation) to (a). In (d), we have the thermofield double state of (a).}
\label{fig:CFTpics}
\end{figure}

From the CFT point of view, the three examples --- half-space (\S \ref{sec:half-space}), cylinder (\S \ref{sec:quench}), and thermofield double (\S \ref{sec:tfd}) --- are essentially the same. %in the previous section all reduce to essentially the same calculation. 
The relevant path-integral geometries are shown in Figure \ref{fig:CFTpics}. In each case, the blue segment(s) show the region whose entanglement entropy we are computing. We insert twist operators $\Phi, \bar{\Phi}$ at the boundaries of these blue regions, and calculate entanglement entropy from their correlator. %The boundaries of the blue regions are the locations where we insert twist operators, $\Phi_n$ and $\bar{\Phi}_n$.

In our first example, the entanglement entropy is calculated via the two-point function of twist operators on the half-plane $\{z = x + iy: \Im(z) = y > 0\}$. The result for a holographic BCFT is
\be
\langle\bar{\Phi}_n(z_1)\Phi_n(z_2)\rangle_\text{UHP}^b = \min\left\{\langle 0|b\rangle^{2(n-1)}\left|\frac{4y_1y_2}{\epsilon^2}\right|^{-d_n}, \left|\frac{z_{12}}{\epsilon}\right|^{-2d_n}\right\}\;,
\label{eq:hol-half-plane}
\ee
for twist scaling dimension $d_n$.
%The results are analytically continued to real time $t = -iy$.
The calculation in \S \ref{sec:tfd} reduces to exactly this correlator under the coordinate transformation
\be
w = f(z) = \frac{1}{z-i/2}-i.
\ee
This maps the boundary of the half-plane to the circle in Figure \ref{fig:CFTpics}d.
Since twists are primary CFT operators (by definition), the correlator in the thermofield double geometry is related to the half-plane correlator by
\be
\langle\bar{\Phi}_n(w_1)\Phi_n(w_2)\rangle_\text{TFD}^b = |w'(z_1)w'(z_2)|^{-d_n}\langle\bar{\Phi}_n(z_1)\Phi_n(z_2)\rangle_\text{UHP}^b.\label{eq:tfd-correlator}
\ee
For a symmetric interval with $w_{1,2} = \pm x_0 + i t$, (\ref{eq:hol-half-plane}) and (\ref{eq:tfd-correlator}) agree with the holographic calculation in the $n\to 1$ limit \cite{Rozali2019}.
%The calculation in the third example reduces to exactly this after we map the circle to the boundary of a half-plane.

The state $|b, \tau_0\rangle$ in \S \ref{sec:quench} takes the CFT on a circle.
This is not related to the half-plane by a conformal transformation, so in principle the correlator of twists must be calculated anew.
However, for a holographic CFT, the dual geometry for small $\tau_0$ is a trivial compacification of the covering space in which we ``unwind" the angular coordinate.
At leading order in large $c$, the entanglement entropy for an interval of fixed length on the circle is then independent of the circle's size.\footnote{This is the phenomenon of large-$N$ volume independence. See, e.g., \cite{Shaghoulian2016}.}

The limit of an infinite circle is the line. This gives the global quench geometry, as in Figure \ref{fig:CFTpics}c.
For small $\tau_0$, this is dual to a planar black hole.
Unlike the cylinder geometry, the global quench is conformally equivalent to the half-plane, under the coordinate transformation
\be
\kappa = \frac{2\tau_0}{\pi}\log z\;.
\ee
The correlator of twists is then
\be
\langle\bar{\Phi}_n(\kappa_1)\Phi_n(\kappa_2)\rangle_\text{GQ}^b = |\kappa'(z_1)\kappa'(z_2)|^{-d_n}\langle\bar{\Phi}_n(z_1)\Phi_n(z_2)\rangle_\text{UHP}^b.\label{eq:quench-correlator}
\ee
Once again, combining this with (\ref{eq:hol-half-plane}) gives the same result as the HRT formula \cite{Cooper2018}.

%For the state $|b, \tau_0 \rangle$ in our second example, the calculation of entanglement entropy involves a Euclidean path integral with two boundaries. In general, this calculation would be unrelated to the other two. However, for a holographic CFT, the dual geometry for small $\tau_0$ is a trivial compacification of a covering space in which the angular coordinate on the circle is taken to be non-compact. This implies that (at leading order in large $N$) the entanglement entropy for an interval on the circle is independent of the size of the circle if the interval size is held fixed and the circle size is increased.\footnote{This is the phenomenon of large $N$ volume independence.} In particular, the result will be the same if the circle size is taken to infinity, giving the path-integral geometry in Figure \ref{fig:CFTpics}c. This strip geometry is related by a conformal transformation to a half-space, and again our calculation becomes the two-point function of twist operators on the half-space. In the next section, we consider this CFT calculation in detail.

\section{Multiple intervals}

The generalization of our BCFT results from a single interval to multiple intervals closely parallels the generalization of the CFT result from two intervals to multiple intervals \cite{Hartman2013}.
We start with the holographic calculation, and then discuss how to obtain the results from the monodromy method in the BCFT.

\subsection{Holographic results for multiple intervals}
\label{sec:mult-hol}

%\Mark{I think that the various types of RT surfaces described below are technically all in the same homology class (since they are required to be homologous to the boundary region). I will change this to the more general ``topology''.}
%\david{I guess this is true if we assume (as in section 3.2) that the boundary is "part of the bulk". I think it is in general a bit subtler, but for our purposes I'm happy to go with "topologies".}

Consider a collection of $k$ disjoint intervals $A = \sqcup_i A_i$, $A_i =
[x_{2i-1},x_{2i}]$, in the vacuum state of a BCFT on the half-space $x \geq 0$, with an associated minimal surface $\mathcal{X}_A$.
A given topology for $\mathcal{X}_A$ geodesically (and without intersection) pairs each endpoint $x_i$ to either (a) another endpoint $x_j$, or (b) the brane.
Morally, we can view the latter as pairing $x_i$ to an image point $x_i^*$ placed on a mirror image of the bulk theory across the brane.

Thus, the possible topologies $\text{top}(\mathcal{X}_A)$ can equally be described by \emph{symmetric} geodesic pairings of $2k$ intervals, of which there are $\binom{2k}{k}$.\footnote{Arbitrary non-intersecting geodesic pairings of $n$ intervals are counted by Catalan numbers $\binom{2k}{2}/(k+1)$. Each yields $k+1$ symmetric pairings on $2k$ intervals, giving our result. We thank Chris Waddell for discussion of this point.} Assuming that the gravity dual theory is described via a purely gravitational ETW brane, so that the local geometry is pure AdS, the two types of geodesics have (regulated) lengths
\[
\frac{\ell_{ij}}{4G_\text{N}} = \frac{c}{3}\log \left(\frac{x_{ij}}{\epsilon}\right), \quad \frac{\ell_{mm^*}}{4G_\text{N}} = \frac{c}{6}\log \left(\frac{2x_{m}}{\epsilon}\right) + \log g_b \;.
\]
Hence, the holographic result is
\begin{align}
    S_A & = \min_{ \text{top}(\mathcal{X}_A)} 
    \frac{1}{4G_\text{N}}\left[ \sum_{(ij)} \ell_{ij} + \sum_{(mm^*)}\ell_{mm^*}\right] \notag \\
    & = \min_{ \text{top}(\mathcal{X}_A)} \left[ \sum_{(ij)} \frac{c}{3}\log \left(\frac{x_{ij}}{\epsilon}\right) + \sum_{(mm^*)} \frac{c}{6}\log \left(\frac{2x_{k}}{\epsilon}\right) + \log g_b\right]\;,
    \label{eq:mult-ee-holo}
\end{align}
where $(ij)$ denotes paired endpoints in the half-space and $(mm^*)$ image-paired endpoints.

As a concrete example, take the interval $A = [x_1, x_2]$.
The explicit expression for holographic entanglement entropy is then
\begin{align*}
  S_A & = \min \left\{\frac{c}{3}\log
    \left(\frac{x_2-x_1}{\epsilon}\right),
    \frac{c}{6}\log\left(\frac{4x_1x_2}{\epsilon^2}\right)+2\log g^b\right\} \\ & = \min\left\{S^{\text{conn}}_A, S^{\text{disc}}_A\right\}\;,
\end{align*}
recovering our results from Section \ref{sec:half-space}.
The calculation is similar in other vacuum geometries. We can also include a boundary-centred interval $[0, x_0]$, which forces at least one image-paired geodesic.\footnote{When $A = [0, x_0] \sqcup A_1 \sqcup \cdots \sqcup A_{k-1}$, $\mathcal{X}_A$ has $\binom{2k+1}{k}$ possible topologies. This can established by similar combinatorics to the non-boundary case.}

\subsection{BCFT calculation for multiple intervals}
\label{sec:bcft-mult}

%\Mark{Can we write this section so that the reader has a sense of where the results above come from without looking at Hartman in detail?}

To calculate the entanglement entropy of $A = \sqcup A_i$ in the BCFT on a half-space, we can simply calculate a correlator of $k$ twist and anti-twist operators on the Euclidean UHP and analytically continue.
We will therefore focus on the UHP calculation.
As above, we can use kinematic doubling to write the correlator as
\begin{align}
\left\langle
\prod_{i=1}^k \Phi_n(z_{2i-1}, \bar{z}_{2i-1})\bar{\Phi}_n(z_{2i}, \bar{z}_{2i})
\right\rangle_\text{UHP}^b = 
\left\langle
\prod_{i=1}^k \Phi_n(z_{2i-1})\bar{\Phi}_n(\bar{z}_{2i-1})\bar{\Phi}_n(z_{2i})\Phi_n(\bar{z}_{2i})
\right\rangle\;.
\end{align}

As in the single interval case, we have some choice about the order in which we perform bulk OPE or BOE expansions of the twist correlator.
We can regard this sequence of choices as a fusion channel $\mathcal{E}$, analogous to the s- and t-channels in the single interval case.
%fusion channel $\mathcal{E}$, corresponding to the sequence of bulk OPE or BOE steps we perform.
% Note that, in the doubled picture, the BOE is an OPE between an operator and its image.
A given fusion channel has a natural expansion in terms of a set of cross-ratios, $\vec{\eta}$, and higher-point conformal blocks:
\begin{align}
    \left\langle
\prod_{i=1}^k \Phi_n(z_{2i-1})\bar{\Phi}_n(\bar{z}_{2i-1})\bar{\Phi}_n(z_{2i})\Phi_n(\bar{z}_{2i})
\right\rangle_\text{UHP}^b = 
N(\vec{\eta})\sum_{\vec{h},\vec{\Delta}} \mathcal{C}^{\mathcal{E},\vec{h},\vec{\Delta}} e^{-\frac{nc}{6}f(\vec{h}, d_n/2nc, \vec{\eta})}\;,
\end{align}
where we have taken the semiclassical limit, and $\mathcal{C}^{\mathcal{E},\vec{h},\vec{\Delta}}$ is a product of OPE and BOE coefficients depending on the internal weights $\vec{h}, \vec{\Delta}$. 
Here $N(\vec \eta)$ is just a standard prefactor. 

Having related the UHP correlator to a chiral correlator, the higher-point blocks can be obtained from the standard monodromy method. 
We briefly summarize this method here, following \cite{Hartman2013}. (We discuss the method in slightly more detail in Appendix \ref{sec:monodromy}.) 
Readers familiar with the monodromy method may freely jump ahead to \eqref{eq:mult-vacuum-1}.

The monodromy method begins with a powerful trick: instead of the desired $2k$-point function, consider a $(2k+1)$-point function, where we have added an additional operator, $\chi_{(1,2)}(z)$, which is taken to be a null descendant of a primary operator $\theta(z)$.\footnote{Strictly speaking, this operator is only guaranteed to exist in Liouville theory. However, as the block is a kinematic object, we expect the form not to depend on whether this operator exists in our theory or not.  }
%The remaining copy of the Virasoro algebra implies that there is a scalar primary $\theta$ with some unphysical null descendant $\chi$. 
%\jamie{I don't understand the statement you are making here. Is the null descendant not just a trick we use to compute the block?}\david{Yes, the field is only guaranteed to exist in Liouville theory where we can compute the block. Maybe I should say that?}
The null operator must decouple and the correlator must vanish.
The vanishing of the correlator is expressed as the differential equation (writing $\chi(z)$ as a differential operator acting on $\theta(z)$) 
\begin{align}
    \Theta''(z) + T(z)\Theta(z) = 0\;,
    \label{eq:fuchsian-1}
\end{align}
where $\Theta(z)$ is the correlator 
\begin{align}
    \Theta(z) = \left\langle\theta(z)
\prod_i \Phi_n(z_{2i-1})\bar{\Phi}_n(\bar{z}_{2i-1})\bar{\Phi}_n(z_{2i})\Phi_n(\bar{z}_{2i})
\right\rangle\;,
\end{align}
and $T(z)$ is 
\begin{align}
    T(z) = \sum_i \left\{\frac{6 h_n/c}{(z-z_i)^2} + \frac{6h_n/c}{(z-\bar{z}_i)^2} + \frac{\partial_{z_i}}{z-z_i}+\frac{\partial_{\bar{z}_i}}{z-\bar{z}_i}\right\}\;.
\end{align}
%Since (\ref{eq:fuchsian-1}) is a second order ODE, it will have two solutions $\Theta^\pm(z)$.

For a given channel $\mathcal{E}$, in an appropriate limit of the cross ratios $\eta \to \eta^{\mathcal{E}}_0$, we expect this to be dominated by the exchange of the lightest possible operator, generally the identity and its descendants. We thus make the ansatz that the correlator is given by
\begin{align}
    %\Theta(z) 
    %\left\langle \prod_i \Phi_n(z_{2i-1}, \bar{z}_{2i-1})\bar{\Phi}_n(z_{2i}, \bar{z}_{2i})
%\right\rangle_\text{UHP}^b
\Theta(z)
\approx \psi(z|z_i,\bar z_i) e^{-\frac{nc}{3}f^\mathcal{E}_0} \, ,
\label{eq:renyi-mult}
\end{align}
to leading order in $c$. 
Here $f^\mathcal{E}_0$ is the semiclassical vacuum block for the original $2k$-point function and $\psi(z|z_i,\bar z_i)$ is thought of as a `wavefunction' for the inserted operator. 
In this case, we can rewrite $T(z)$ %the stress tensor as
\begin{align}
    T(z) = \sum_i \left\{\frac{6 h_n/c}{(z-z_i)^2} + \frac{6h_n/c}{(z-\bar{z}_i)^2} - \frac{c_i}{z-z_i}-\frac{\bar{c}_i}{z-\bar{z}_i}\right\}\;,
    \label{eq:T2}
\end{align}
where the $c_i$ are \emph{accessory parameters}:
\begin{align}\label{eq:accessory-param}
    c_i = \frac{\partial f^{\mathcal{E}}_0}{\partial z_i}, \quad \bar{c}_i = \frac{\partial f^{\mathcal{E}}_0}{\partial \bar{z}_i} = \overline{c_i}\;.
\end{align}

If we know the accessory parameters, we can integrate \eqref{eq:accessory-param} to find the block $f^{\mathcal{E}}_0$.
To determine these parameters, the monodromy method then uses the fact that a solution of the differential equation must have monodromies around any set of points that is consistent with the corresponding operator being exchanged in the block. 
This constraint can be used to fix the accessory parameters. 
In general, this cannot be done analytically.  
However, it is possible to find the parameters explicitly for twist operators in the $n\to 1$ limit, when we can break the problem down into a sum of independent monodromy constraints.

\begin{figure}[t]
  \centering
  \includegraphics[scale=0.5]{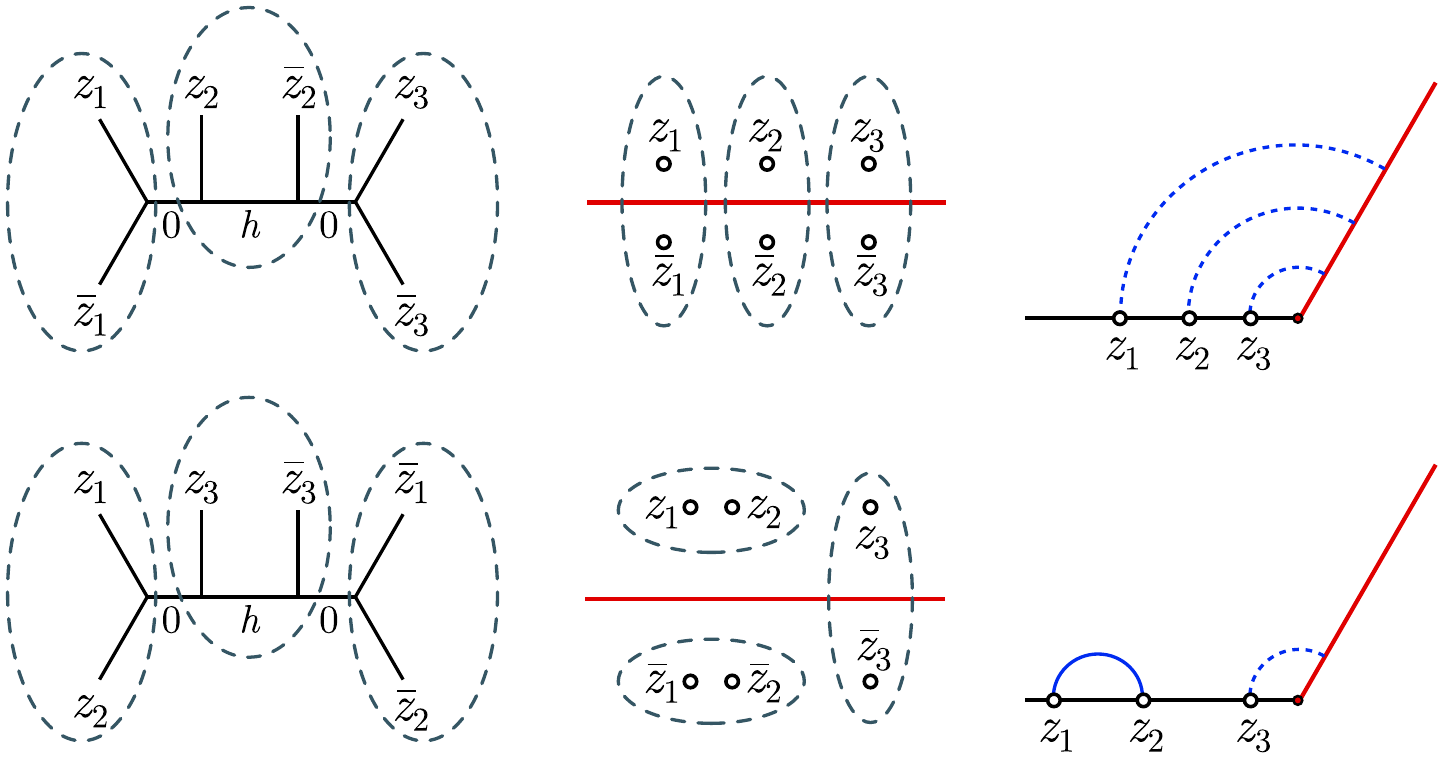}
  \caption{\emph{Left.} Vacuum exchange in two different channels for $k = 3$ twists on the UHP.
    Trivial cycles cut through identities. \emph{Middle.} The monodromy cycles to be trivialized in the doubled picture of the BCFT. \emph{Right.} The corresponding RT topologies in the bulk with an ETW brane.}
\label{fig:monodromy}
\end{figure}

Solving the monodromy constraints and integrating the accessory parameters near $n=1$, one finds
\begin{align}
    f^{\mathcal{E}}_0 = \sum_{(ij)} 12\alpha \log |z_{ij}|^2 + \sum_{(mm^*)} 12\alpha \log z_{mm^*} + O(\alpha^2)\;.
    \label{eq:mult-vacuum-1}
\end{align}
Here, the channel $\mathcal{E}$ pairs some twists to anti-twists on the same half-plane, and some twists to their images on the opposite half-plane.
We have denoted the pairs by $(ij)$ and $(mm^*)$ respectively, and note that (as expected from the CFT case \cite{Hartman2013}) channels biject with the topologies of Section \ref{sec:mult-hol}, so we can view $\mathcal{E} \in \text{top}(\mathcal{X}_A)$.
We illustrate the correspondence between channels, trivial cycles, and the bulk RT surfaces in Figure \ref{fig:monodromy}.

To calculate the entanglement entropy, we also need to compute $\mathcal{C}^{\mathcal{E}}_0$.
This is easily done, since the OPE coefficients for vacuum exchange are always unity, while the BOE always gives the one-point function of twists (\ref{BPhi}).
If there are $M$ image pairs $(mm^*)$, we have
\begin{align}
    \mathcal{C}^{\mathcal{E}}_0 = [g_b^{1-n}]^{M} = g_b^{-12\alpha M}\;.
\end{align}
We can recover the factors of $\epsilon$ from the one-point functions (\ref{eq:Z_n}) and (\ref{BPhi}).
From (\ref{eq:renyi-mult}), the entanglement entropy in the limit $\eta \to \eta^{\mathcal{E}}_0$ is then
\begin{align}
    S_A & = \lim_{\alpha \to 0}  \left(\frac{c}{36\alpha}f^{\mathcal{E}}_0 - \frac{1}{12\alpha}\log \mathcal{C}^{\mathcal{E}}_0\right) \notag \\ & = \sum_{(ij)} \frac{c}{3} \log \left(\frac{|z_{ij}|}{\epsilon}\right) + \sum_{(mm^*)} \frac{c}{6} \log \left(\frac{z_{mm^*}}{\epsilon}\right) + \log g_b\;.
    \label{eq:mult-cft-ee-1}
\end{align}
The corrections to (\ref{eq:mult-vacuum}) are in $\alpha^2$ and not in $z_{ij}$.
It follows that in \emph{finite} regions around $\eta^{\mathcal{E}}_0$, expression (\ref{eq:mult-cft-ee-1}) is the full entanglement entropy to leading order in $c$.

If we make the assumption of vacuum block dominance as in Section \ref{sec:renyi}, we can upgrade (\ref{eq:mult-cft-ee-1}) to precisely reproduce (\ref{eq:mult-ee-holo}):
\begin{align}
    S_A & = \min_{\text{top}(\mathcal{X}_A)}\left[ \sum_{(ij)} \frac{c}{3} \log \left(\frac{|z_{ij}|}{\epsilon}\right) + \sum_{(mm^*)} \frac{c}{6} \log \left(\frac{z_{mm^*}}{\epsilon}\right) + \log g_b\right]\;.
    \label{eq:mult-cft-ee-2}
\end{align}
This follows because vacuum dominance in a channel $\mathcal{E}$ implies the vacuum contribution is larger in other channels.
Thus, we have a derivation of the full RT formula in a BCFT dual to AdS with an ETW brane, to the same level of generality as the CFT case \cite{Hartman2013}.

%As expected, a choice of fusion channel $\mathcal{E}$ corresponds to a choice of topology for the RT surfaces described in Section \ref{sec:mult-hol}.

\section{Replica calculation in the gravity picture}

In this paper, we have seen how phase transitions in holographic BCFT entanglement entropies, originally understood using the HRT formula, can be understood directly in the BCFT via the exchange of dominance between bulk and boundary channels in the two-point function of twist operators. The twist operator correlation function calculates the R\'{e}nyi entropies as the partition function for the BCFT on a replica manifold. Our calculation indicates that the R\'{e}nyi entropies themselves, or the replica partition functions, also have a phase transition.\footnote{Note that the parameter values at which this happens depends on the replica index.} It is interesting to understand the gravitational origin of these transitions. Here, we use the fact that the CFT partition function on the replica manifold should be equal to the partition function of the gravity theory in which the spacetime geoemtries are constrained to be asymptotically AdS, with boundary geometry equal to the replica manifold.

In \cite{Rozali2019}, inspired by \cite{Almheiri2019qdq,Penington2019kki}, it was argued that the transitions in the gravitational path integral arise because we can have various topologies for the ETW brane whose boundary is the disconnected set of $n$ CFT boundaries, where $n$ is the replica index. In the saddle-point approximation (which gives the leading contribution to the partition function in the $1/c$ expansion), the logarithm of the CFT partition function is equal to the gravitational action for the bulk configuration with least action. The transition occurs when the gravitation configuration with least action changes as the parameters specifying the interval are varied. 

In this section, we will confirm these expectations in an explicit example where the geometries can be understood in detail. We consider the R\'{e}nyi entropy for a single interval $[x_1,x_2]$ in the vacuum state of a BCFT on a half-line $x > 0$. We assume that the BCFT and the chosen boundary condition correspond to a dual gravitational theory that has an effective three-dimensional description as gravity with a negative cosmological constant and an gravitational ETW brane with zero tension. In this simple case, the dual geometries that we need can be obtained very simply from the geometries that contribute to the calculation of the R\'{e}nyi entropies for a pair of intervals $[-x_2,-x_1] \cup [x_1,x_2]$ in the vacuum state of our parent CFT on a real line. These geometries have a $\mathbb{Z}_2$ symmetry corresponding to the transformation $x \to -x$ in the CFT that exchanges the intervals.

It follows from symmetry that the bulk surface fixed by this transformation has zero extrinsic curvature.\footnote{To see this, recall that the extrinsic curvature can be defined as the Lie derivative of the hypersurface metric with respect to a normal vector $n$ to the hypersurface. This switches sign under $n \to -n$, but the extrinsic curvature is invariant under this operation because of the symmetry.} Thus, if we consider a new geometry defined as half of the previous geometry, with a zero-tension ETW brane replacing the $\mathbb{Z}_2$ symmetric surface, the new geometry will satisfy the gravitational equations in the bulk and at the ETW brane. The new geometry is thus a valid saddle-point geometry for the BCFT R\'{e}nyi entropy calculation. It is plausible that all of the geometries we need can be obtained in this way.

\begin{figure}[t]
  \centering
  \includegraphics[scale=0.7]{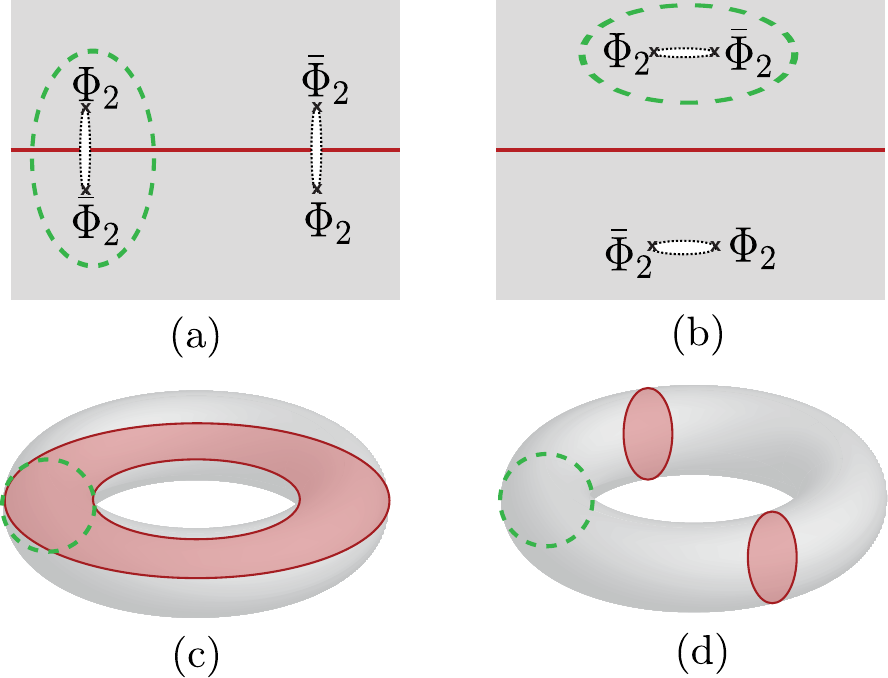}
  \caption{CFT and bulk depictions of the path integral for computing the second R\'{e}nyi entropy of two intervals. We indicate a $\mathbb{Z}_2$ identification, whose fixed-point locus is drawn as a red line, that relates this to a BCFT whose boundary is the red line. (a) When the twist operators are near the $\mathbb{Z}_2$-line (in red), the corresponding bulk solution will fill in the cycle indicated by the green dashed line. The $\mathbb{Z}_2$-line and the bulk contractible cycle intersect each other. (b) When the twist operators are far from the $\mathbb{Z}_2$-line, the bulk contractible cycle (green) is now homologous to the $\mathbb{Z}_2$-line on each sheet. (c) The corresponding bulk solution when the twist operators are near the $\mathbb{Z}_2$-line. The bulk $\mathbb{Z}_2$-surface connects the two $\mathbb{Z}_2$-lines on each sheet. (d) The bulk solution when the twist operators are far from the $\mathbb{Z}_2$-line. There is now a disconnected bulk $\mathbb{Z}_2$-surface for each $\mathbb{Z}_2$-line on the boundary. }
\label{fig:bulk-second-renyi-z2}
\end{figure}

Happily, the geometries for the two-interval calculation have already been discussed in detail by Faulkner in \cite{Faulkner2013}. There, it was understood that the phase transition in R\'{e}nyi entropies indeed arises from a transition in the lowest-action gravitational solution. We can check that under this transition, the hypersurface fixed by the $\mathbb{Z}_2$ symmetry changes topology, becoming connected when the intervals are close to each other. 
We illustrate this for the case of the second Renyi entropy in Figure \ref{fig:bulk-second-renyi-z2}. 
Thus, the ETW brane in the gravity version of the BCFT calculation takes on a replica wormhole geometry in the gravitational saddle that computes the entanglement entropy of an interval close to the boundary. 

\begin{figure}[t]
  \centering
  \includegraphics[scale=0.2]{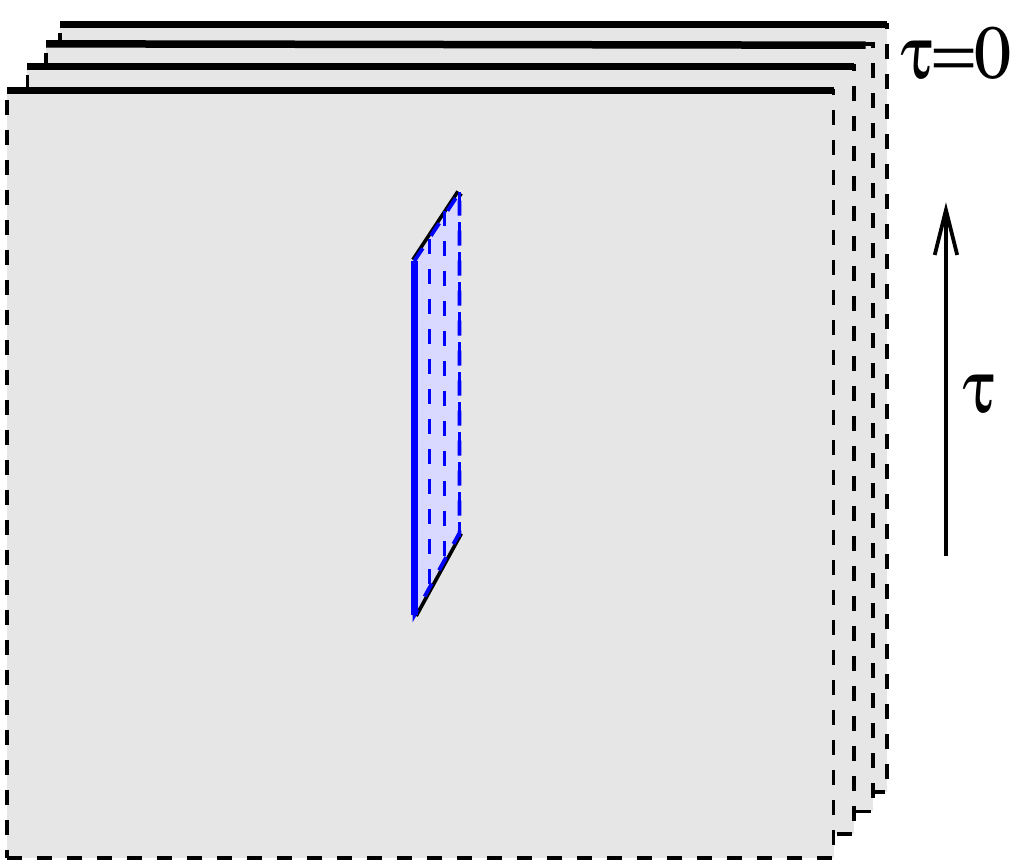}
  \caption{Entangled state of $n$ CFTs produced by inserting twist and anti-twist operators into the Euclidean path integral for the vacuum state. As the branch cut is moved toward $\tau=0$, the entanglement between the CFTs becomes large, and the $\tau=0$ spatial slice of the dual geometry becomes a connected multi-boundary wormhole. In the original picture, this surface is the ETW brane geometry in the gravitational calculation of R\'{e}nyi entropy for an interval close to the BCFT boundary.}
\label{fig:nCFT}
\end{figure}

There is an alternative intuitive picture for the transition in the ETW-brane topology that leads to the phase transition in R\'{e}nyi entropies. Consider again the two-interval calculation, but now reinterpret the $x$ coordinate as a Euclidean time coordinate $\tau$. In this case, the CFT path-integral on the $\tau < 0$ part of the replica manifold can be interpreted as a Euclidean path integral that creates a state of $n$ CFTs. This is illustrated in Figure \ref{fig:nCFT}. The $\mathbb{Z}_2$ symmetric surface in the dual gravitational geometry is the $\tau =0$ surface that gives the initial data for the time-symmetric Lorentzian geometry dual to this state. 

In the limit where our interval moves toward $\tau = -\infty$, the path integral will give the vacuum state of $n$ CFTs, so the dual geometry is $n$ copies of AdS and the $\mathbb{Z}_2$ symmetric surface has $n$ disconnected components. On the other hand, as the interval moves closer to $\tau=0$, the path integral produces a state with more and more entanglement between the $n$ CFTs.\footnote{For example, in the $n=2$ case, we get a path integral similar to the one that produces the thermofield double state, and moving the interval closer to $\tau=0$ corresponds to increasing the temperature.}\footnote{In the limit where the interval hits $\tau=0$, we have a state where the left half of $\text{CFT}_k$ is connected to the right half of $\text{CFT}_{k+1}$, so the entanglement of the system formed from the left and right half of $\text{CFT}_k$ is infinite.} At some point, we have a phase transition similar to the Hawking-Page transition, where the $\tau=0$ spatial slice in the dual geometry becomes connected and the Lorentzian geometry dual to our state is a multi-boundary wormhole. In our BCFT application, this $\tau=0$ spatial slice becomes the ETW brane geometry, so we see that the Euclidean wormholes in the replica calculation can be directly related to the usual appearance of wormholes in the gravity dual of highly-entangled states of holographic systems. 

\section{Discussion}

Starting with the vacuum state of a 1+1-dimensional CFT, the geometry of a putative bulk dual is fixed by symmetry to be AdS$_3$, up to internal dimensions.
We can also fix AdS$_3$ using the RT formula: it is the unique bulk geometry whose minimal surfaces correctly reproduce the universal result for the entanglement entropy of a single interval.
The RT formula makes \emph{non-universal} predictions for two or more intervals, so we can go in the other direction and determine the class of \emph{holographic CFTs} which reproduce these non-universal gravitational results.
As shown in \cite{Hartman2013}, vacuum block dominance guarantees that the twist-antitwist correlators used to calculate entanglement entropy agree with the holographic value for any number of intervals.
Vacuum dominance places explicit constraints on the spectrum and OPE coefficients of a holographic CFT.

% Entanglement entropy can in general be calculated via the replica trick, and in 2d CFTs, this reduces to a local field theory calculation, since $n$-R\'{e}nyi entropies are obtained from twist-antitwist correlators.
% The gravitational prediction is precisely the vacuum exchange diagram for these correlators, so holographic CFTs obey vacuum block dominance 
% Gravitational physics 
% A necessary and sufficient condition is vacuum dominance \cite{Hartman2013}, i.e. dominance of vacuum exchange in correlators, which has implications for the spectrum and OPE coefficients of the CFT.
% In fact, these conditions reproduce the holographic predictions for an \emph{arbitrary} boundary region, consisting of many disjoint intervals.

The logic for a CFT with boundary is similar.
Symmetry, or the universal result for the entanglement entropy of a boundary-centred interval, restricts us to a class of $\text{SO}(1,2)$-invariant geometries. These can include warping in the bulk and compact internal dimensions as before.
The simplest bulk geometry is a portion of AdS$_3$ cut off by an ETW brane with purely gravitational couplings \cite{Takayanagi2011}, though we emphasize this is not the most general bulk dual consistent with ground-state symmetry.

In this paper, we have taken the next step of transforming non-universal gravitational predictions from these geometries with a purely gravitational ETW brane into a constraint on holographic BCFTs.
To match the holographic predictions for a non-centred interval, or indeed any number of intervals, vacuum dominance in both the BCFT bulk and boundary channels is necessary and sufficient.
From a kinematic perspective, this follows from the doubling trick and the remaining copy of the Virasoro algebra.
But the implications for the BCFT spectrum and OPE coefficients are more subtle.
We have made some precise statements above, but expect there is more juice to be squeezed from this particular lemon.
For instance, it might be possible to finesse the spectral constraints along the lines of \cite{Hartman2014}, though the CFT machinery required is potentially quite different. It would also be interesting to investigate the additional constraints that arise from assuming not only that the BCFT calculations reproduce the leading ${O(c)}$ entropies, but also that the subleading corrections to the entropies are order $c^0$ as we expect from a conventional gravitational theory with a semiclassical expansion.

We have argued that the expression (\ref{disc}) for the small $\eta$ entanglement entropy is universal  in holographic CFTs (assuming that the RT surface is disconnected in some interval $[0,\eta_d]$ as in the simple model), so the constraints associated with vacuum block dominance for an interval around $\eta=0$ should be expected to hold much more generally, for any holographic BCFT with a disconnected RT surface phase at small $\eta$. It seems plausible that any holographic BCFT should have such a phase, though it would be interesting to find a direct argument. 

Our results have several interesting consequences and applications.
First, they put the AdS/BCFT proposal of \cite{Takayanagi2011} on firmer microscopic footing, exhibiting explicit conditions on a BCFT under which a locally AdS geometry with a purely gravitational ETW brane captures the microscopic ground-state entanglement entropies. The gravity calculations allow the RT surface to end on an ETW brane, so our results also confirm this aspect of Takayanagi's proposal.\footnote{From the lower-dimensional perspective, this is a modification of the usual homology condition, though no modification is required if we take the higher-dimensional perspective that the ETW brane represents a smooth part of the full bulk geometry.} 

%Although it remains somewhat puzzling from the standpoint of entanglement wedge reconstruction, our results confirm that the  is modified so that RT surfaces can end on the brane, as proposed in \cite{Cooper2018}.
%(If the brane is part of the bulk manifold from the higher-dimensional perspective, no modification is required at all.)

Our work also has direct applications to the physics of black holes. For black holes dual to CFT states prepared by a Euclidean path integral on the cylinder with conformally invariant boundary conditions, the phase transition in entanglement entropy for a non-centred interval leads to a period of Lorentzian time where boundary observers with access to suitably large boundary regions can see behind the horizon \cite{Cooper2018}.

Similarly, for the thermofield-double state of a BCFT on a half-line, the phase transition in entanglement entropy corresponds to a transition in bulk entanglement wedge to include part of the black hole interior \cite{Rozali2019}.
Treating this as a model of a black hole in equilibrium with its Hawking radiation, after this transition part of the interior is reconstructable from the radiation. The brane is therefore playing a similar role to the ``quantum extremal islands" which restore unitarity of the Page curve \cite{Penington:2019npb,Almheiri2019b,Almheiri2019a}.
We have argued above that the phase transition in entanglement entropy can be directly related to topological changes in bulk replica wormholes, and it is of obvious interest to explore this connection further.

% \david{Any other future directions?}
% \jamie{Will think if there is more to add, but I think it reads really well so far.}

\subsection*{Acknowledgments}

We would like to thank Tarek Anous, Thomas Hartman, Eliot Hijano, Alex May, Dominik Neuenfeld, and Chris Waddell for useful discussions. DW is supported by an International Doctoral Fellowship from the University of British Columbia.  MVR is supported by the Simons Foundation via the It From Qubit Collaboration and a Simons Investigator Award. This work is supported in part by the Natural Sciences and Engineering Research Council of Canada.

\appendix

\section{BCFT two-point functions from Virasoro conformal blocks. }

In this appendix, we briefly review the structure of four-point functions of chiral operators and their expansion in terms of Virasoro conformal blocks, and then argue that the same objects form the building blocks of two-point functions in boundary conformal field theories.

\subsubsection*{Chiral four-point functions and conformal blocks}

In a 2D CFT, for operators $\phi_i$ with chiral dimensions $h_i$, the global conformal symmetry implies that the four-point function takes the form
\be
\label{chiral}
\langle \phi_1(z_1) \phi_2(z_2) \phi_3(z_3) \phi_4(z_4) \rangle =  \left(\frac{z_{24}}{z_{14}}\right)^{h_{1}-h_{2}}\left(\frac{z_{14}}{z_{13}}\right)^{h _3- h_4} \frac{\eta^{h_1 + h_2}}{(z_{12})^{h_{1}+h_{2}} (z_{34})^{ h_{3}+ h_{4}}} F(\eta)
\ee
where $z_{ij} = z_i - z_j$ and $F$ is some function of the single cross-ratio $\eta = z_{12} z_{34} / (z_{13} z_{24})$. We can define $F$ as
\be
F(\eta) = \lim_{z_\infty \to \infty} (-1)^{h_1+h_2+h_3+h_4} z_\infty^{2 h_4} \langle \phi_1(0) \phi_2(\eta) \phi_3(1) \phi_4(z_\infty) \rangle \; .
\ee
We can express $\mathcal{F}$ in terms of the OPE data for the CFT and a standard set of functions by expanding the products $\phi_1(z_1) \phi_2(z_2)$ and $\phi_3(z_3) \phi_4(z_4)$ using (\ref{ope}). In this case, the four point function reduces to a sum of two-point functions of intermediate operators,
\be
\label{blocks}
F(\eta) = \sum_i C_{12}^i C_{34}^i \mathcal{F}(c, h ; [h_1,h_2,h_3,h_4] | \eta) \; ,
\ee
The conformal blocks $\mathcal{F}(c, h ; [h_1,h_2,h_3,h_4] | \eta)$ are specific functions which depend only on the central charge, the dimensions $h_i$ of the external operators, and the ``internal'' dimension $h$. These give the contribution to the four-point function from a primary operator of weight $h$ and all of its Virasoro descendants. The block has a simple behavior in the limit $\eta \to 0$, where we have 
\be
\mathcal{F}(c, h ; [h_1,h_2,h_3,h_4] | \eta \to 0) \sim \eta^{h - h_1 - h_2}
\ee

\subsubsection*{BCFT two-point function}

We now consider the two-point function of bulk operators in a BCFT defined on the upper-half-plane
\begin{equation}
\label{App2pt}
   \uavg{ \Ocal_1(z_1, \bar{z}_1) \Ocal_2(z_2, \bar{z}_2)} \, .
\end{equation}
As we discussed in section 2, this  has the kinematics of a chiral four-point function. We will show this somewhat more carefully here, and see that we can expand the two-point function in either a bulk channel or a boundary channel in terms of the chiral conformal blocks defined above.

\subsubsection*{One-point functions for scalar Virasoro descendants}

To begin, it will be useful to  compute one-point functions for scalar global primaries that are themselves Virasoro descendants. 
Consider, in particular, a Virasoro primary operator $\Ocal_{h,\bar h}(z, \bar{z})$ and state $|h,\bar{h}\rangle$ for the CFT on $S^1$ associated to it by the state-operator correspondence. We denote an operator $\Ocal^{\alpha,\beta}_{h,\bar h}(0)$ which creates the Virasoro descendants of this state $V^h_\alpha  \bar V^{\bar h}_\beta |h,\bar{h}\rangle$. Here, $V^h_\alpha$ and $V^{\bar h}_\beta$ are polynomials in $L_{-n}$ and $\bar{L}_{-n}$ respectively chosen so that these states give an orthonormal basis of the Verma module:\footnote{Recall that the conjugation operation used to define the dual operator at infinity is an inversion in radial quantization, and so the operator is rescaled by the conformal transformation. We keep the rescaling implicit.} 
\begin{equation}
\label{ortho}
    \avg{\Ocal^{\alpha',\beta' }_{h,\bar h}(\infty) \Ocal^{\alpha,\beta}_{h,\bar h}(0)}=\bracket{h,\bar{h}}{V^{h,\dagger}_{\alpha^\prime} \bar V^{\bar h,\dagger}_{\beta^\prime} V^h_\alpha  \bar V^{\bar h}_\beta }{h,\bar{h}} = \delta_{\alpha\alpha^\prime}  \delta_{\beta\beta^\prime} \, .
\end{equation}
We can re-express the same operators in terms of local operators $ \Ocal^{\alpha_i,\beta_i}_{h,\bar h}(z_i,\bar z_i)$ at arbitrary points $z_1,z_2$  by the use of a global conformal transformation that maps $(\infty,0)$ to $(z_1,z_2)$. 
We then have 
\be
\label{ortho2}
\avg{ \Ocal^{\alpha_1,\beta_1}_{h,\bar h}(z_1,\bar{z}_1)  \Ocal^{\alpha_2,\beta_2}_{h,\bar h}(z_2,\bar{z}_2)} = \delta_{\alpha_1\alpha_2}  \delta_{\beta_1\beta_2} \, .
\ee
Note that the form of each local descendant operator depends explicitly on both points $z_1,z_2$, and not just implicitly on one point through the local primary. 
These operators are only orthogonal precisely at the points $z_1,z_2$ (and form an orthogonal basis of operators in the `North-South' quantization between these two points).

Next, we require a somewhat more refined version of the doubling trick. We have seen that a correlator 
\be
\label{UHP2}
\uavg{{\cal O}_{h_1 \bar h_1}(z_1,\bar{z}_1) \cdots {\cal O}_{h_n \bar h_n}(z_n,\bar{z}_n) } 
\ee
of bulk CFT operators ${\cal O}_{h_k \bar h_k}$ with conformal weights $(h_k,\bar{h}_k)$ is constrained to have the same functional form as chiral CFT correlators 
\be
\label{ChiCorr2}
\langle {\cal O}_{h_1}(z_1) \cdots {\cal O}_{h_n}(z_n) {{\cal O}}_{\bar h_1}(\bar{z}_1) \cdots {{\cal O}}_{\bar h_n}(\bar{z}_n) \rangle
\ee
Similarly, a correlator of descendants
\be
\label{UHPD}
\uavg{{\cal O}^{\alpha_1, \beta_1}_{h_1 \bar h_1}(z_1,\bar{z}_1) \cdots {\cal O}^{\alpha_n, \beta_n}_{h_n \bar h_n}(z_n,\bar{z}_n) } 
\ee
takes the same functional form as 
\be
\label{ChiCorrD}
\langle {\cal O}^{\alpha_1}_{h_1}(z_1) \cdots {\cal O}^{\alpha_n}_{h_n}(z_n) {{\cal O}}^{\beta_1}_{\bar h_1}(\bar{z}_1) \cdots {{\cal O}}^{\beta_n}_{\bar h_n}(\bar{z}_n) \rangle \, .
\ee

Then, taking $\Ocal_{h}(z)$ to be a primary operator in some CFT such that\footnote{Note that it's not necessary for such a CFT to exist, since we are only making statements about kinematics.}
\bea
\uavg{\Ocal_{h, h}(z,\bar z)} &=& \mathcal{A}^b_h \avg{\Ocal_{ h}(\bar{z}) \Ocal_{h}(z)} \cr
&=& {\mathcal{A}^b_h \over |z - \bar{z}|^{2h} }
\eea
we have that
\bea
   \uavg{\Ocal^{\alpha,\beta}_{h, h}(z,\bar z)} &=& \mathcal{A}^b_h \avg{\Ocal^{\alpha}_{h}(z) \Ocal^{\beta}_{h}(\bar{z})   } \cr
   &=& \delta_{\alpha \beta} {\mathcal{A}^b_h}
   \label{eq:virasoro-descendant-one-point}
\eea
where here the descendant indices are labeling the orthogonal basis of states for the pair of points $z,
\bar z$.

\subsubsection*{Bulk channel expression for the two-point function}

We can now derive a bulk-channel expression for the two-point function (\ref{App2pt}). First we will use the bulk state-operator map (bulk OPE) to insert a complete set of bulk states (in this `North-South' quantization between $z_3$ and $\bar z_3$)
\begin{align}
    \uavg{ \Ocal_1(z_1, \bar{z}_1) \Ocal_2(z_2, \bar{z}_2)} = \sum_{i, \alpha, \beta} {\langle \Ocal_1(z_1, \bar{z}_1) \Ocal_2(z_2, \bar{z}_2) \Ocal_i^{\alpha,\beta}(\bar z_3, z_3) \rangle  \uavg{ \Ocal_i^{\alpha,\beta}(z_3,\bar z_3) }}  \,.
\end{align}
Using the form of the boundary one-point function \eqref{eq:virasoro-descendant-one-point}, we can rewrite this as
\begin{align}
    \uavg{ \Ocal_1(z_1, \bar{z}_1) \Ocal_2(z_2, \bar{z}_2)} = \sum_{\{i|h_i = \bar{h}_i\}, \alpha, \beta} \hat\Ccal^i_{12} \Acal^b_i {\langle \Ocal_{h_1}(z_1) \Ocal_{h_2}(z_2) \Ocal_{h_i}^{\alpha}(\bar z_3) \rangle  \langle \Ocal_{h_i}^{\alpha}(z_3) \Ocal_{\bar h_1}(\bar z_1) \Ocal_{\bar h_2}(\bar z_2) \rangle}  \, ,
\end{align}
where we have pulled out the dynamical information in the OPE coefficients and expectation values.  The three-point functions, as written, are now purely kinematic, i.e. they represent the functional dependence of such a three-point function where the overall coefficient is taken to be one. 
Each sum over Virasoro descendants now can be seen to give a standard chiral Virasoro conformal block $\vblock{h}{h_1}{h_2}{\bar h_1}{\bar h_2}{z}$, so that the two-point function can be expanded in this \emph{bulk channel} as 
\begin{align}\label{eq:bulk-two-point-bulk-channel}
    \uavg{ \Ocal_1(z_1, \bar{z}_1) \Ocal_2(z_2, \bar{z}_2)} = &\left(\frac{z_{21^*}}{z_{11^*}}\right)^{h_{1}-h_{2}}\left(\frac{z_{11^*}}{z_{12^*}}\right)^{\bar h _{2}-\bar h_{1}} \frac{z^{h_1 + h_2}}{(z_{12})^{h_{1}+h_{2}} (z_{2^*1^*})^{\bar h_{1}+\bar h_{2}}} \nonumber \\
    &\times \sum_{i} \hat\Ccal^i_{12} \Acal^b_i \vblock{h_i}{h_1}{h_2}{\bar h_1}{\bar h_2}{z}\,,
\end{align}
and where we have written the conformal block in terms of the cross-ratio
\begin{equation}
    z = \frac{z_{12}z_{2^*1^*} }{z_{12^*}z_{21^*}} \, .
\end{equation}

\subsubsection*{Boundary channel expression for the two-point function}

We can similarly expand the two-point function in the \emph{boundary channel}. 
Here we insert a complete set of states corresponding to the expansion of the bulk operators in terms of the boundary operator expansion. 
The boundary state-operator mapping gives a complete set of states in terms of boundary operators which appear in representations of the surviving diagonal Virasoro symmetry. 
We thus insert a complete set of orthonormal states of the form 
\begin{equation}
    \uavg{ \Ocal_1(z_1, \bar{z}_1) \Ocal_2(z_2, \bar{z}_2)} = \sum_{I, \alpha} \uavg{ \Ocal_1(z_1, \bar{z}_1)  \tilde V^{\hat h_I}_\alpha  | \hat h_I } \uavg{ \hat h_I | \tilde V^{\hat h_I \dagger}_\alpha \Ocal_2(z_2, \bar{z}_2) } \,.
\end{equation}
%\jamie{OR:}
%\begin{equation}
%    \uavg{ \Ocal_1(z_1, \bar{z}_1) \Ocal_2(z_2, \bar{z}_2)} = \sum_{I, \alpha} \uavg{ \Ocal_1(z_1, \bar{z}_1)  \Ocal^{\alpha}_I(0) } \uavg{ \Ocal^{\alpha \, }_I(\infty) \Ocal_2(z_2, \bar{z}_2) } \,.
%\end{equation}
Using the doubling trick to account for the representation of bulk operators under the boundary Virasoro operators, we can rewrite this as
\begin{equation}
    \uavg{ \Ocal_1(z_1, \bar{z}_1) \Ocal_2(z_2, \bar{z}_2)} = \sum_{I, \alpha} \Bcal^b_{1h}\Bcal^b_{2h} \avg{ \Ocal_{h_1}(z_1) \Ocal_{\bar h_1}({z}^*_1)  V_\alpha^{\hat h_I}  | \hat h_I } \avg{\hat h_I | V^{\hat h_I \dagger}_\alpha \Ocal_{h_2}(z_2) \Ocal_{\bar h_2}({z}^*_2) }\,,
\end{equation}
%\jamie{OR:}
%\begin{equation}
%    \uavg{ \Ocal_1(z_1, \bar{z}_1) \Ocal_2(z_2, \bar{z}_2)} = \sum_{I, \alpha} \Bcal^b_{1h}\Bcal^b_{2h} \avg{ \Ocal_{h_1}(z_1) \Ocal_{\bar h_1}({z}^*_1)  \Ocal^{\alpha}_I(0) } \avg{ \Ocal^{\alpha }_I(\infty) \Ocal_{h_2}(z_2) \Ocal_{\bar h_2}({z}^*_2) }\,,
%\end{equation}
where we have pulled out the dynamical information in the coefficients. 
The remaining three-point functions, as written, are now purely kinematic. 
Again we recognize that this sum over Virasoro descendants is the standard  bulk chiral Virasoro conformal block $\vblock{h}{h_1}{\bar h_1}{h_2}{\bar h_2}{\eta}$, giving
\begin{align}\label{eq:bulk-two-point-boundary-channel}
    \uavg{ \Ocal_1(z_1, \bar{z}_1) \Ocal_2(z_2, \bar{z}_2)} = &\left(\frac{z_{1^*2}}{z_{12}}\right)^{h_{1}-\bar h_{1}}\left(\frac{z_{12}}{z_{12^*}}\right)^{\bar h_{2}-h_{2}} \frac{\eta^{h_1 +\bar h_1}}{(z_{11^*})^{h_{1}+\bar h_{1}} (z_{22^*})^{h_{2}+\bar h_{2}}} \nonumber \\
    &\times \sum_{I} \Bcal^b_{1I}\Bcal^b_{2I}  \vblock{h}{h_1}{\bar h_1}{h_2}{\bar h_2}{\eta}
\end{align}
where we have used the cross-ratio 
\begin{equation}
    \eta = 1-z  \, .
\end{equation}

\section{Boundary operator expansion for twist operators}\label{sec:app-twist-boe-coeff}

In this section, we relate the boundary operator expansion of the twist operator $\Phi_n$ in an $n$-copy BCFT to $n$-point functions of boundary operators in the original BCFT. Our discussion here is directly parallel to the discussion in Section 4 of \cite{Perlmutter2014} on contributions to the OPE coefficients of CFT twist operators. 

Via radial quantization, a twist operator inserted at $z$ into an $n$-copy BCFT can be understood to give rise to some entangled state of this $n$-copy BCFT on an interval. By the state-operator correspondence, the same state can be obtained by the insertion of some operator at the origin. A basis of boundary operators for the $n$-copy BCFT may be written as ${\cal O}_{I_1} \otimes \cdots \otimes {\cal O}_{I_n}$, where ${\cal O}_{I}$ are a basis of boundary operators in the original BCFT. Thus, we can write that
\be
\Phi_n(x+iy) = \sum_{\{I_k\}} {1 \over |2y|^{d_n - \sum_k \Delta_{I_k}}} B^{\Phi_n}_{I_1 \cdots I_n} {\cal O}_{I_1} \otimes \cdots \otimes {\cal O}_{I_n}(x)\;.
\ee 
When the operators ${\cal O}_{I_1}$ are primary, the coefficient $B^{\Phi_n}_{I_1 \cdots I_n}$ can be defined according to (\ref{boundarybulk}) via the bulk-boundary two-point function as
\beas
B^{\Phi_n}_{I_1 \cdots I_n} &=& 2^{d_n - \sum_k \Delta_{I_k}} \langle \Phi_n(z=i)  {\cal O}_{I_1} \otimes \cdots \otimes {\cal O}_{I_n}(0) \rangle \cr
&=& 2^{d_n - \sum_k \Delta_{I_k}} \langle \Phi_n(i) \rangle {\langle \Phi_n(i)  {\cal O}_{I_1} \otimes \cdots \otimes {\cal O}_{I_n}(0) \rangle \over \langle \Phi_n(i) \rangle } \cr
&=& 2^{- \sum_k \Delta_{I_k}} \epsilon^{d_n} g_b^{1-n} {\langle \Phi_n(i)  {\cal O}_{I_1} \otimes \cdots \otimes {\cal O}_{I_n}(0) \rangle \over \langle \Phi_n(i) \rangle }\;.
\eeas 
To compute the ratio of correlators in the last line, consider the conformal transformation
\be
w(z) = i \left({(z + i)^n + (z - i)^n \over (z + i)^n - (z - i)^n }\right)\;.
\ee
This takes the UHP to the $n$-sheeted UHP associated with the insertion of our twist operator. The points
\be
x_k \equiv \cot \left( \pi {2k-1 \over 2 n} \right) \qquad k = 1, \dots, n
\ee
map to the origin on the various sheets. By this conformal transformation, we have that
\be
{\langle \Phi_n(i)  {\cal O}_{I_1} \otimes \cdots \otimes {\cal O}_{I_n}(0) \rangle \over \langle \Phi_n(i) \rangle } = \prod_k \left({\D w \over \D z}(x_k)\right)^{-\Delta_{I_k}} \left\langle \prod_k {\cal O}_{I_k}(x_k) \right\rangle
\ee 
For the points $x_k$ where $w(z) = 0$, we have that 
\be
{\D w \over \D z}(x_k) = {n \over x_k^2 + 1} = n \sin^2 \left(\pi {2k-1 \over 2 n} \right)
\ee
Combining everything, we have that 
\be
B^{\Phi_n}_{I_1 \cdots I_n} = 2^{ - \sum_k \Delta_{I_k}} \epsilon^{d_n} g_b^{1-n}\prod_k \left[] n \sin^2 \left(\pi {2k-1 \over 2 n} \right) \right]^{-\Delta_{I_k}} \left \langle \prod_k {\cal O}_{I_k}(x_k) \right\rangle\;.
\ee
It is useful to note that the explicit dependence on $c$ appears as a universal prefactor,
\be
B^{\Phi_n}_{I_1 \cdots I_n} = \epsilon^{d_n} g_b^{1-n} \bar B^{\Phi_n}_{I_1 \cdots I_n}\;.
\ee
For $n=2$, we see that the correlator vanishes unless $I_1 = I_2$, and we have that
\be
\bar B^{\Phi_2}_{I I} =  {1 \over 16^{\Delta_I}}\;.
\ee
For $n=3$, we have that 
\be
\bar B^{\Phi_3}_{I J K} = {C_{IJK} \over 3^{{3 \over 2}(\Delta_I + \Delta_J + \Delta_K)}}\;,
\ee
where we have used the standard result for a CFT three-point function.

\section{Monodromy method}
\label{sec:monodromy}
Here we continue the discussion of accessory parameters from  \eqref{eq:accessory-param} in the main text to give a more complete description of the calculation of the semiclassical blocks.

There are only $2k-3$ independent accessory parameters, since global $\text{SL}(2,\mathbb{R})$ invariance imposes three (real) constraints.
Explicitly, these constraints are
\begin{align*}  
\sum_{i} \Re(c_{i}) = \sum_{i} \Re \left(c_{i} z_{i} - \frac{6h_n}{c}\right) = \sum_{i}\Re\left(c_{i} z_{i}^2 -
  \frac{12h_nz_{i}}{c}\right) = 0\;,
\end{align*}
the real part of the usual $\text{SL}(2,\mathbb{C})$ constraints.

If we know the accessory parameters, we can integrate to find the block $f^{\mathcal{E}}_0$.
To determine these parameters, we transport a pair of solutions $\vec{\Theta}(z) = [\Theta^+(z),\Theta^-(z)]^T$ around a point $z_c$ where an OPE or BOE is to be performed.
The null decoupling equation (\ref{eq:fuchsian-1}) applied to the three-point function implies that the $2\times 2$ monodromy matrix $M$ performing the transport, $\vec{\Theta}(z) \mapsto M\vec{\Theta}(z)$, %gives %\jamie{Is it correct to call this trivial? }, with
gives\footnote{To see this, we suppose the leading term in $\Theta(z) \sim (z-z_c)^\kappa$. Plugging this into (\ref{eq:fuchsian-1}), we find that $\kappa(\kappa - 1) = -6h_n/c$, with two solutions $\kappa_\pm$.
These pick up factors $e^{2\pi i \kappa_\pm}$ after traversing a loop $z = z_c + \epsilon e^{i\theta}$, leading to (\ref{eq:mono}). See \cite{Hartman2013} for details.}
\begin{align}
    \mbox{tr} M = -2\cos(\pi \Lambda_c), \quad \Lambda_c = \sqrt{1-\frac{24 h_n}{c}}.
    \label{eq:mono}
\end{align}
The number of independent monodromies to tune equals the number of internal primaries, $2k-3$,\footnote{An exchange channel $\mathcal{E}$ is a cubic tree with $2k$ leaves and $2k-2$ internal nodes in the doubled picture. The total number of edges is one less than the number of nodes, $E = 4k-3$, and hence the number of internal edges is $E - 2k = 2k -3$.} so we have the right number of monodromy constraints to fix our accessory parameters $c_i$.
%We can then integrate to find the vacuum block $f^{\mathcal{E}}_0$ in the regime it dominates.

In general, we cannot analytically solve for the accessory parameters.
Luckily, however, it is possible to find them explicitly for twist operators in the $n\to 1$ limit.
As above, we define $\alpha = (n-1)/12$.
Entanglement entropy is obtained from R\'{e}nyi entropies in the limit $\alpha \to 0$, and since $h_n = c(n+1)\alpha/2n = c\alpha \to 0$ in this limit, the function (\ref{eq:T2}) vanishes away from the singular points $z_i, \bar{z}_i$.
As a result, the equation (\ref{eq:fuchsian-1}) decouples into a sum of independent monodromy equations, depending on which cycles the channel $\mathcal{E}$ trivializes.

To illustrate, suppose $\mathcal{E}$ involves a pairing between twists $\Phi_n(z_i)$ and $\bar{\Phi}_n(z_j)$.
We must choose the accessory parameters to make the monodromy around $z_i, z_j$ trivial.
Since this decouples from the other problems as $\alpha \to 0$, we can simply focus on the contribution
\begin{align}
T_{ij}(z) & = \frac{6h_n}{c}\left[\frac{1}{(z-z_i)^2} + \frac{1}{(z-z_j)^2}\right] - \frac{c_i}{z-z_i}-\frac{c_j}{z-z_j} + \text{c.c.} \notag \\
& = 6\alpha\left[\frac{1}{(z-z_i)^2} + \frac{1}{(z-z_j)^2} - \frac{2}{z_j(z-z_i)}\right] - \frac{c_i}{z-z_i}+\frac{c_i z_i}{z_j(z-z_j)}+ \text{c.c.}\;,
\label{eq:Tij}
\end{align}
where ``c.c" stands for complex conjugate terms, and in (\ref{eq:Tij}), we used the  constraint $\Re(c_i z_i + c_j z_j) = 6\alpha$.
To obtain a trivial monodromy around $z_i, z_j$ (and the image cycle enclosing $\bar{z}_i, \bar{z}_j$), it is sufficient for $T_{ij}(z)$ to be regular at infinity.
This is equivalent to the sum of residues at simple poles vanishing, and hence
\begin{align}
    %-\frac{12\alpha}{z_j}- c_i + \frac{c_iz_i}{z_j} = 0 \quad \Longrightarrow \quad 
    c_i + \bar{c}_i = \frac{12\alpha}{|z_{ij}|^2} + O(\alpha^2)\;,
    \label{eq:accessory}
\end{align}
where $O(\alpha^2)$ corrections arise because the equations only strictly decouple for $\alpha = 0$.
The calculation is analogous for a twist paired with its image, but the contribution $T_{mm^*}(z)$ involves only two insertions at $z_m$ and $\bar{z}_m$.

If we integrate the accessory parameters defined in (\ref{eq:accessory}) (and the image-paired counterparts), we find
\begin{align}
    f^{\mathcal{E}}_0 = \sum_{(ij)} 12\alpha \log |z_{ij}|^2 + \sum_{(mm^*)} 12\alpha \log z_{mm^*} + O(\alpha^2)\;
    \label{eq:mult-vacuum}
\end{align}
as required.

\bibliographystyle{jhep}
\bibliography{BCFTDraft}

\providecommand{\href}[2]{#2}\begingroup\raggedright\begin{thebibliography}{10}

\bibitem{Ryu2006a}
S.~Ryu and T.~Takayanagi, \emph{{Aspects of Holographic Entanglement Entropy}},
  \href{https://doi.org/10.1088/1126-6708/2006/08/045}{\emph{JHEP} {\bfseries
  08} (2006) 045} [\href{https://arxiv.org/abs/hep-th/0605073}{{\ttfamily
  hep-th/0605073}}].

\bibitem{Karch2000}
A.~Karch and L.~Randall, \emph{{Open and closed string interpretation of SUSY
  CFT's on branes with boundaries}},
  \href{https://doi.org/10.1088/1126-6708/2001/06/063}{\emph{JHEP} {\bfseries
  06} (2001) 063} [\href{https://arxiv.org/abs/hep-th/0105132}{{\ttfamily
  hep-th/0105132}}].

\bibitem{Takayanagi2011}
T.~Takayanagi, \emph{{Holographic Dual of BCFT}},
  \href{https://doi.org/10.1103/PhysRevLett.107.101602}{\emph{Phys. Rev. Lett.}
  {\bfseries 107} (2011) 101602}
  [\href{https://arxiv.org/abs/1105.5165}{{\ttfamily 1105.5165}}].

\bibitem{Anous2016}
T.~Anous, T.~Hartman, A.~Rovai and J.~Sonner, \emph{{Black Hole Collapse in the
  1/c Expansion}}, \href{https://doi.org/10.1007/JHEP07(2016)123}{\emph{JHEP}
  {\bfseries 07} (2016) 123}
  [\href{https://arxiv.org/abs/1603.04856}{{\ttfamily 1603.04856}}].

\bibitem{Cooper2018}
S.~Cooper, M.~Rozali, B.~Swingle, M.~Van~Raamsdonk, C.~Waddell and D.~Wakeham,
  \emph{{Black Hole Microstate Cosmology}},
  \href{https://arxiv.org/abs/1810.10601}{{\ttfamily 1810.10601}}.

\bibitem{Rozali2019}
M.~Rozali, J.~Sully, M.~Van~Raamsdonk, C.~Waddell and D.~Wakeham,
  \emph{{Information radiation in BCFT models of black holes}},
  \href{https://arxiv.org/abs/1910.12836}{{\ttfamily 1910.12836}}.

\bibitem{Penington:2019npb}
G.~Penington, \emph{{Entanglement Wedge Reconstruction and the Information
  Paradox}},  \href{https://arxiv.org/abs/1905.08255}{{\ttfamily 1905.08255}}.

\bibitem{Almheiri2019b}
A.~Almheiri, N.~Engelhardt, D.~Marolf and H.~Maxfield, \emph{{The entropy of
  bulk quantum fields and the entanglement wedge of an evaporating black
  hole}},  \href{https://arxiv.org/abs/1905.08762}{{\ttfamily 1905.08762}}.

\bibitem{Almheiri2019a}
A.~Almheiri, R.~Mahajan, J.~Maldacena and Y.~Zhao, \emph{{The Page curve of
  Hawking radiation from semiclassical geometry}},
  \href{https://arxiv.org/abs/1908.10996}{{\ttfamily 1908.10996}}.

\bibitem{Hartman2013}
T.~Hartman, \emph{{Entanglement Entropy at Large Central Charge}},
  \href{https://arxiv.org/abs/1303.6955}{{\ttfamily 1303.6955}}.

\bibitem{Faulkner2013}
T.~Faulkner, \emph{{The Entanglement Renyi Entropies of Disjoint Intervals in
  AdS/CFT}},  \href{https://arxiv.org/abs/1303.7221}{{\ttfamily 1303.7221}}.

\bibitem{Almheiri2019qdq}
A.~Almheiri, T.~Hartman, J.~Maldacena, E.~Shaghoulian and A.~Tajdini,
  \emph{{Replica Wormholes and the Entropy of Hawking Radiation}},
  \href{https://arxiv.org/abs/1911.12333}{{\ttfamily 1911.12333}}.

\bibitem{Penington2019kki}
G.~Penington, S.~H. Shenker, D.~Stanford and Z.~Yang, \emph{{Replica wormholes
  and the black hole interior}},
  \href{https://arxiv.org/abs/1911.11977}{{\ttfamily 1911.11977}}.

\bibitem{McAvity1995}
D.~M. McAvity and H.~Osborn, \emph{{Conformal field theories near a boundary in
  general dimensions}},
  \href{https://doi.org/10.1016/0550-3213(95)00476-9}{\emph{Nuclear Physics,
  Section B} {\bfseries 455} (1995) 522}
  [\href{https://arxiv.org/abs/9505127}{{\ttfamily 9505127}}].

\bibitem{DiFrancesco1997}
P.~Di~Francesco, P.~Mathieu and D.~Senechal, \emph{{Conformal Field Theory}},
  Graduate Texts in Contemporary Physics. Springer-Verlag, New York, 1997,
  \href{https://doi.org/10.1007/978-1-4612-2256-9}{10.1007/978-1-4612-2256-9}.

\bibitem{Cardy2004}
J.~L. Cardy, \emph{{Boundary conformal field theory}},
  \href{https://arxiv.org/abs/hep-th/0411189}{{\ttfamily hep-th/0411189}}.

\bibitem{Calabrese2009}
P.~Calabrese and J.~Cardy, \emph{{Entanglement entropy and conformal field
  theory}}, \href{https://doi.org/10.1088/1751-8113/42/50/504005}{\emph{J.
  Phys.} {\bfseries A42} (2009) 504005}
  [\href{https://arxiv.org/abs/0905.4013}{{\ttfamily 0905.4013}}].

\bibitem{Liendo2013}
P.~Liendo, L.~Rastelli and B.~C. {Van Rees}, \emph{{The bootstrap program for
  boundary CFT d}},
  \href{https://doi.org/10.1007/JHEP07(2013)113}{\emph{Journal of High Energy
  Physics} {\bfseries 2013} (2013) }
  [\href{https://arxiv.org/abs/1210.4258}{{\ttfamily 1210.4258}}].

\bibitem{Calabrese2016}
P.~Calabrese and J.~Cardy, \emph{{Quantum quenches in 1 + 1 dimensional
  conformal field theories}},
  \href{https://doi.org/10.1088/1742-5468/2016/06/064003}{\emph{J. Stat. Mech.}
  {\bfseries 1606} (2016) 064003}
  [\href{https://arxiv.org/abs/1603.02889}{{\ttfamily 1603.02889}}].

\bibitem{Recknagel:2013uja}
A.~Recknagel and V.~Schomerus, \emph{{Boundary Conformal Field Theory and the
  Worldsheet Approach to D-Branes}}, Cambridge Monographs on Mathematical
  Physics. Cambridge University Press, 2013,
  \href{https://doi.org/10.1017/CBO9780511806476}{10.1017/CBO9780511806476}.

\bibitem{Karch:2000gx}
A.~Karch and L.~Randall, \emph{{Open and closed string interpretation of SUSY
  CFT's on branes with boundaries}},
  \href{https://doi.org/10.1088/1126-6708/2001/06/063}{\emph{JHEP} {\bfseries
  06} (2001) 063} [\href{https://arxiv.org/abs/hep-th/0105132}{{\ttfamily
  hep-th/0105132}}].

\bibitem{Takayanagi:2011zk}
T.~Takayanagi, \emph{{Holographic Dual of BCFT}},
  \href{https://doi.org/10.1103/PhysRevLett.107.101602}{\emph{Phys. Rev. Lett.}
  {\bfseries 107} (2011) 101602}
  [\href{https://arxiv.org/abs/1105.5165}{{\ttfamily 1105.5165}}].

\bibitem{Fujita:2011fp}
M.~Fujita, T.~Takayanagi and E.~Tonni, \emph{{Aspects of AdS/BCFT}},
  \href{https://doi.org/10.1007/JHEP11(2011)043}{\emph{JHEP} {\bfseries 11}
  (2011) 043} [\href{https://arxiv.org/abs/1108.5152}{{\ttfamily 1108.5152}}].

\bibitem{Astaneh:2017ghi}
A.~Faraji~Astaneh and S.~N. Solodukhin, \emph{{Holographic calculation of
  boundary terms in conformal anomaly}},
  \href{https://doi.org/10.1016/j.physletb.2017.03.026}{\emph{Phys. Lett.}
  {\bfseries B769} (2017) 25}
  [\href{https://arxiv.org/abs/1702.00566}{{\ttfamily 1702.00566}}].

\bibitem{Chiodaroli:2011fn}
M.~Chiodaroli, E.~D'Hoker and M.~Gutperle, \emph{{Simple Holographic Duals to
  Boundary CFTs}}, \href{https://doi.org/10.1007/JHEP02(2012)005}{\emph{JHEP}
  {\bfseries 02} (2012) 005} [\href{https://arxiv.org/abs/1111.6912}{{\ttfamily
  1111.6912}}].

\bibitem{Chiodaroli:2012vc}
M.~Chiodaroli, E.~D'Hoker and M.~Gutperle, \emph{{Holographic duals of Boundary
  CFTs}}, \href{https://doi.org/10.1007/JHEP07(2012)177}{\emph{JHEP} {\bfseries
  07} (2012) 177} [\href{https://arxiv.org/abs/1205.5303}{{\ttfamily
  1205.5303}}].

\bibitem{DHoker:2007zhm}
E.~D'Hoker, J.~Estes and M.~Gutperle, \emph{{Exact half-BPS Type IIB interface
  solutions. I. Local solution and supersymmetric Janus}},
  \href{https://doi.org/10.1088/1126-6708/2007/06/021}{\emph{JHEP} {\bfseries
  06} (2007) 021} [\href{https://arxiv.org/abs/0705.0022}{{\ttfamily
  0705.0022}}].

\bibitem{DHoker:2007hhe}
E.~D'Hoker, J.~Estes and M.~Gutperle, \emph{{Exact half-BPS Type IIB interface
  solutions. II. Flux solutions and multi-Janus}},
  \href{https://doi.org/10.1088/1126-6708/2007/06/022}{\emph{JHEP} {\bfseries
  06} (2007) 022} [\href{https://arxiv.org/abs/0705.0024}{{\ttfamily
  0705.0024}}].

\bibitem{Aharony:2011yc}
O.~Aharony, L.~Berdichevsky, M.~Berkooz and I.~Shamir, \emph{{Near-horizon
  solutions for D3-branes ending on 5-branes}},
  \href{https://doi.org/10.1103/PhysRevD.84.126003}{\emph{Phys. Rev.}
  {\bfseries D84} (2011) 126003}
  [\href{https://arxiv.org/abs/1106.1870}{{\ttfamily 1106.1870}}].

\bibitem{Assel:2011xz}
B.~Assel, C.~Bachas, J.~Estes and J.~Gomis, \emph{{Holographic Duals of D=3 N=4
  Superconformal Field Theories}},
  \href{https://doi.org/10.1007/JHEP08(2011)087}{\emph{JHEP} {\bfseries 08}
  (2011) 087} [\href{https://arxiv.org/abs/1106.4253}{{\ttfamily 1106.4253}}].

\bibitem{Almheiri2018}
A.~Almheiri, A.~Mousatov and M.~Shyani, \emph{{Escaping the Interiors of Pure
  Boundary-State Black Holes}},
  \href{https://arxiv.org/abs/1803.04434}{{\ttfamily 1803.04434}}.

\bibitem{Cardy:2007mb}
J.~L. Cardy, O.~A. Castro-Alvaredo and B.~Doyon, \emph{{Form factors of
  branch-point twist fields in quantum integrable models and entanglement
  entropy}}, \href{https://doi.org/10.1007/s10955-007-9422-x}{\emph{J. Statist.
  Phys.} {\bfseries 130} (2008) 129}
  [\href{https://arxiv.org/abs/0706.3384}{{\ttfamily 0706.3384}}].

\bibitem{Cardy2016}
J.~Cardy and E.~Tonni, \emph{{Entanglement hamiltonians in two-dimensional
  conformal field theory}},
  \href{https://doi.org/10.1088/1742-5468/2016/12/123103}{\emph{J. Stat. Mech.}
  {\bfseries 1612} (2016) 123103}
  [\href{https://arxiv.org/abs/1608.01283}{{\ttfamily 1608.01283}}].

\bibitem{Belavin1984}
A.~A. Belavin, A.~M. Polyakov and A.~B. Zamolodchikov, \emph{{Infinite
  Conformal Symmetry in Two-Dimensional Quantum Field Theory}},
  \href{https://doi.org/10.1016/0550-3213(84)90052-X}{\emph{Nucl. Phys.}
  {\bfseries B241} (1984) 333}.

\bibitem{Dorn1994}
H.~Dorn and H.~J. Otto, \emph{{Two and three point functions in Liouville
  theory}}, \href{https://doi.org/10.1016/0550-3213(94)00352-1}{\emph{Nucl.
  Phys.} {\bfseries B429} (1994) 375}
  [\href{https://arxiv.org/abs/hep-th/9403141}{{\ttfamily hep-th/9403141}}].

\bibitem{Zamolodchikov1995}
A.~B. Zamolodchikov and A.~B. Zamolodchikov, \emph{{Structure constants and
  conformal bootstrap in Liouville field theory}},
  \href{https://doi.org/10.1016/0550-3213(96)00351-3}{\emph{Nucl. Phys.}
  {\bfseries B477} (1996) 577}
  [\href{https://arxiv.org/abs/hep-th/9506136}{{\ttfamily hep-th/9506136}}].

\bibitem{Harlow2011a}
D.~Harlow, J.~Maltz and E.~Witten, \emph{{Analytic Continuation of Liouville
  Theory}}, \href{https://doi.org/10.1007/JHEP12(2011)071}{\emph{JHEP}
  {\bfseries 12} (2011) 071} [\href{https://arxiv.org/abs/1108.4417}{{\ttfamily
  1108.4417}}].

\bibitem{Zamolodchikov1987}
A.~B. Zamolodchikov, \emph{Conformal symmetry in two-dimensional space:
  Recursion representation of conformal block},
  \href{https://doi.org/10.1007/BF01022967}{\emph{Theoretical and Mathematical
  Physics} {\bfseries 73} (1987) 1088}.

\bibitem{Perlmutter2014}
E.~Perlmutter, \emph{{Comments on Renyi entropy in AdS$_3$/CFT$_2$}},
  \href{https://doi.org/10.1007/JHEP05(2014)052}{\emph{JHEP} {\bfseries 05}
  (2014) 052} [\href{https://arxiv.org/abs/1312.5740}{{\ttfamily 1312.5740}}].

\bibitem{Pappadopulo:2012jk}
D.~Pappadopulo, S.~Rychkov, J.~Espin and R.~Rattazzi, \emph{{OPE Convergence in
  Conformal Field Theory}},
  \href{https://doi.org/10.1103/PhysRevD.86.105043}{\emph{Phys. Rev.}
  {\bfseries D86} (2012) 105043}
  [\href{https://arxiv.org/abs/1208.6449}{{\ttfamily 1208.6449}}].

\bibitem{Chen:2013kpa}
B.~Chen and J.-J. Zhang, \emph{{On short interval expansion of Rényi
  entropy}}, \href{https://doi.org/10.1007/JHEP11(2013)164}{\emph{JHEP}
  {\bfseries 11} (2013) 164} [\href{https://arxiv.org/abs/1309.5453}{{\ttfamily
  1309.5453}}].

\bibitem{Calabrese:2010he}
P.~Calabrese, J.~Cardy and E.~Tonni, \emph{{Entanglement entropy of two
  disjoint intervals in conformal field theory II}},
  \href{https://doi.org/10.1088/1742-5468/2011/01/P01021}{\emph{J. Stat. Mech.}
  {\bfseries 1101} (2011) P01021}
  [\href{https://arxiv.org/abs/1011.5482}{{\ttfamily 1011.5482}}].

\bibitem{doi:10.1112/plms/s2-17.1.75}
G.~H. Hardy and S.~Ramanujan, \emph{Asymptotic formulaæ in combinatory
  analysis}, \href{https://doi.org/10.1112/plms/s2-17.1.75}{\emph{Proceedings
  of the London Mathematical Society} {\bfseries s2-17} (1918) 75}
  [\href{https://arxiv.org/abs/https://londmathsoc.onlinelibrary.wiley.com/doi/pdf/10.1112/plms/s2-17.1.75}{{\ttfamily
  https://londmathsoc.onlinelibrary.wiley.com/doi/pdf/10.1112/plms/s2-17.1.75}}].

\bibitem{Shaghoulian2016}
E.~Shaghoulian, \emph{{Emergent gravity from Eguchi-Kawai reduction}},
  \href{https://doi.org/10.1007/JHEP03(2017)011}{\emph{JHEP} {\bfseries 03}
  (2017) 011} [\href{https://arxiv.org/abs/1611.04189}{{\ttfamily
  1611.04189}}].

\bibitem{Hartman2014}
T.~Hartman, C.~A. Keller and B.~Stoica, \emph{{Universal Spectrum of 2d
  Conformal Field Theory in the Large c Limit}},
  \href{https://doi.org/10.1007/JHEP09(2014)118}{\emph{JHEP} {\bfseries 09}
  (2014) 118} [\href{https://arxiv.org/abs/1405.5137}{{\ttfamily 1405.5137}}].

\end{thebibliography}\endgroup

\end{document}